\documentclass[journal=jacsat,manuscript=article]{achemso}

\usepackage{chemformula} 
\usepackage[T1]{fontenc} 
\usepackage{multirow}
\usepackage{chemformula}

\usepackage{amsmath}
\usepackage{xr}
\usepackage{rotating}
\usepackage{afterpage}
\makeatletter
\newcommand*{\addFileDependency}[1]{
\typeout{(#1)}
\@addtofilelist{#1}
\IfFileExists{#1}{}{\typeout{No file #1.}}
}\makeatother
\newcommand*{\myexternaldocument}[1]{
\externaldocument{#1}%
\addFileDependency{#1.tex}%
\addFileDependency{#1.aux}%
}
\myexternaldocument{suppInfo}

\author{Anseong Park}
\affiliation[SNU]
{School of Chemical and Biological Engineering and Institute of Chemical Processes, Seoul National University}
\author{Jaeyune Ryu}
\affiliation[SNU]
{School of Chemical and Biological Engineering and Institute of Chemical Processes, Seoul National University}
\author{Won Bo Lee}
\email{wblee@snu.ac.kr}
\affiliation[SNU]
{School of Chemical and Biological Engineering and Institute of Chemical Processes, Seoul National University}

\title[]
  {Ionic Liquid Molecular Dynamics Simulation with Machine Learning Force Fields: DPMD and MACE}

\abbreviations{MD, DFT, TFSI, PYR$_{14}$, MLFF}

\begin{document}

\begin{abstract}
Machine learning force fields (MLFFs) are gaining attention as an alternative to classical force fields (FFs) by using deep learning models trained on density functional theory (DFT) data to improve interatomic potential accuracy. In this study, we develop and apply MLFFs for ionic liquids (ILs), specifically PYR$_{14}$BF$4$ and LiTFSI/PYR$_{14}$TFSI, using two different MLFF frameworks: DeePMD (DPMD) and MACE. We find that high-quality training datasets are crucial, especially including both equilibrated (EQ) and non-equilibrated (nEQ) structures, to build reliable MLFFs. Both DPMD and MACE MLFFs show good accuracy in force and energy predictions, but MACE performs better in predicting IL density and diffusion. We also analyze molecular configurations from our trained MACE MLFF and notice differences compared to pre-trained MACE models like MPA-0 and OMAT-0. Our results suggest that careful dataset preparation and fine-tuning are necessary to obtain reliable MLFF-based MD simulations for ILs.
\end{abstract}

\section{Introduction}
Density functional theory (DFT) simulation and molecular dynamics (MD) simulation have made tremendous progress over the past several decades since the first emergence of molecular simulation in 1940s. DFT and MD simulation works as a trade-offs, where DFT simulation focuses on angstrom femtosecond scale chemical details lays on electron-level interactions, while MD simulation focuses on nanometer nanosecond scale structural and dynamical properties.
However, MD simulation possesses two fundamental limitations. First is that due to their point-charge method, it is rationally impractical to develop a force field (FF) for a large heavy metal atom due to their large electron cloud and polarization effects. Second, they must rely on pre-existing FF parameters which were calculated through n DFT laborious simulations and FF parameters are transferable only when they were obtained from the exact same DFT method. Whenever newly advanced DFT method has newly developed, one has to start from the scratch to make the FF transferable, making it praticle in real research. The fact that numerous researches now are based on OPLS-AA FF, developed in 1996, shows the limitations of the current MD simulation.\cite{shi2024molecular,robertson2022development, kaminski2001evaluation} 
Under the current circumstance, machine learning force fields (MLFFs) or machine learned interatomic potentials (MLIPs) has been developed as a next generation FF as a alternative to traditional MDFFs. While detailed implementations may differ, the MLFF approximate the potential energy surface (PES) of the given structure from pre-trained machine learning models (train data from DFT calculations) and calculates forces, virials, and energies.\cite{kang2020large}

Various MLFF building open software packages have been developed: DeePMD, DPGEN, MACE, Nequip, UF3, SevenNet, Allegro and etc, where each approach presents its own strengths and weaknesses.\cite{wang2018deepmd,zhang2020dp,batatia2022mace,batzner20223,xie2023ultra,park2024scalable,ibayashi2023allegro} Though MLFF technology was developed relatively recently, various studies have already been conducted and proved that the well built MLFF MD simulation represents the accurate forces and energy values obtained from ab initio simulation.\cite{zhang2021phase,jiang2021accurate,owen2024complexity} Introduction of pre-trained foundational models, such as MPA-0 and OMAT-0 provided by MACE, enables us to train (fine-tune) a new MLFF model with time and computational cost efficiently.\cite{batatia2023foundation} Moreover, pre-trained foundational models itself could also achieve more than 98\% accuracy without fine-tuning.\cite{nwachukwu2022microstructure} However, unlike the guaranteed-systematic building process of classic MDFFs, the process building a MLFF largely involves the expertise of a researcher, specifically, preparation of high quality training datasets and well defined hyper parameters are crucial.\cite{ni2024pre,tropsha2024integrating} Though the researches validate their newly developed MLFFs through ab initio validation datasets, one has to mention that the quality of validation datasets significant affects the tested MLFF validation score. Also, while the force and energy deviation of the MLFF validation tests may stay within the tolerance range, whether the MLFF's PES covers the whole possible structures is indeed remains uncertain.
This study aims to discuss three key topics. First is, preparation of fine quality training datasets investigated through binary ionic liquid (IL) system PYR$_{14}$BF$_4$. MLFF was built with DPMD and the MD simulation results was compared with classic FF and polarizable FF MD simulation.\cite{doherty2017revisiting,bedrov2019molecular} We found that not only the equilibrated (EQ) dataset, but also the non-equilibrated (nEQ) dataset was critical to improve the performance of MLFF. Second is, building and evaluation of MLFF built by MACE with ternary IL system LiTFSI/PYR$_{14}$TFSI. The results were compared with experimental data and showed that the MLFF achieved reasonable accuracy compared to classic and polarizable MD simulation. Lastly, evaluation of single molecule configuration built by MLFF is discussed. Our trained MLFF, which resembles the molecular configuration of polarizable MD simulation, showed different molecular configuration compared to pre-trained MACE models MPA-0 and OMAT-0. 
We believe that though MLFF MD simulation may run stable without issues such as molecular entanglement or fragmentation, it may produce incorrect molecular structures or interaction patterns without fine-tuning of pre-trained MLFF, or fine quality training datasets. Overall study raises a question that does the newly built MLFF really "correct?"

\section{Method}
\textbf{non-Pol MD} GROMACS simulation tool with CL\&P force field was 
used for non-polarizable classic MD simulation.\cite{abraham2015gromacs,canongia2012cl,goloviznina2019transferable} For all the simulated systems, initial configuration were constructed using PACKMOL.\cite{martinez2009packmol} The cutoff distance for short-range electrostatic interactions and van der Waals interactions was set to 1 nm.\cite{cunkui2010Effect} The particle mesh Ewald (PMF) method was used for long-range interactions further than the cutoff distance.\cite{york1993effect} Integration time step of 2fs was used for both Nos$\acute{e}$-Hoover thermostat and Parrinello-Rahman pressure coupling throughout the simulation.\cite{nose1984unified,hoover1985canonical,parrinello1981polymorphic} \\
\textbf{Pol MD} WMI-MD with APPLE$\&$P FF was used to represent polarization effects of the molecules.\cite{bedrov2019molecular} Ewald summation method was used for long-range electrostatic interactions of charge-charge and charge-induced dipole interactions, a reaction-field approximation tempering function was used for induced dipole-induced dipole interactions, while the many body polarization tolerance was set to 0.001.\cite{toukmaji1996ewald} Three different time scales for the integrator were adopted to accelerate the simulation speed; short (0.5 fs) for bonds, bends and improper torsions, medium (1.5 fs) for torsions and short-range nonbonded interactions (within 6.5 \AA), and long (3.0 fs) for electrostatic interactions and full-range nonbonded interactions.\cite{lippert2013accurate}\\
\textbf{DFT/AIMD} For VASP DFT calculation, two functional were tested while other parameters were remained the same; generalized gradient approximation (GGA) adopting Perdew-Burke-Ernzerhof (PBE) exchange-correlation functional and meta-GGA adopting strongly constrained and appropriately normed (SCAN) exchange-correlation functional.\cite{kresse1993ab,perdew1996generalized,hammer1999improved,sun2015strongly} All conducted systems were consists of 10 ion-pairs (PYR$^{14}$BF$_4$ or LiTFSI/PYR$_{14}$TFSI), where the box size ranges from 1.4 nm to 1.6 nm. The projector-augmented-wave (PAW) method was used with an energy cutoff of the plane-wave basis set 450 eV, EDIFF (global break condition for the electronic SC-loop) was set to 1e$^{-6}$ eV, sigma (with of the smearing) was set to 0.05, and DFT-D3 method was utilized to describe accurate vdW interactions.\cite{kresse1999ultrasoft} Gamma-point calculation was performed (only the gamma point in the Brillouin zone is used for sampling the k-points) to accelerate the production speed.
NVT AIMD simulation using VASP were performed controlling the box size and initial equilibrium to embrace all the possible molar configurations. The initials for equilibrated systems were prepared using GROMACS NVT MD simulation run, while initials for non-equilibrium systems undergo very short NVT MD simulation or taken directly from PACKMOL's randomly generated structure. The details of DFT/AIMD datasets are shown in Table~\ref{tbl:AIMDdataset}\\
\textbf{MLFF} In this study, DPGEN and MACE were used build the MLFF, LAMMPS was used to run DPGEN MLFF MD simulation and OpenMM was used to run MACE MLFF MD simulation.\cite{thompson2022lammps,eastman2017openmm} \\ 
For DPGEN MLFF building process, the cut-off radius was set to 10 \AA. The size of embedding and fitting nets were set to [25, 50, 100] and [240, 240, 240]. The learning rate started from 1.0 x 10$^{-3}$ and exponentially decayed to 3.5 x 10$^{-8}$ after 1 x 10$^6$ steps, while final model was trained with 2 x 10$^6$ steps. The start and end of prefactor of energy, force, and virial was set to (0.02, 1), (1000, 1), and (0.02, 1), respectively. Two-body descriptor (se\_e2\_a) was used to represent the local atomic environment and four models were trained utilizing Adam stochastic gradient descent method with its random seed for initializing model parameters. The models were then tested through LAMMPS MLFF simulation with pre-prepared training-sets. Each training-set simulation is consists of 10 ion-pairs and the simulation was conducted under 450K NPT, 1 x 10$^5$ steps with trajectory frequency 1 x 10$^3$. The low and high boundary of force deviation was set to 0.05 eV/\AA and 1.5 eV/\AA, respectively, to select candidates for inclusion to next MLFF training dataset.\\
For MACE MLFF bilding process, ScaleShiftMace model was used with hidden irrpes 16x0e+4x1e, batch size of 4, and the cut-off radius was set to 6 \AA. The maximum number of epochs was set to 100 with two stage loss functions, where the weights of energy and forces for the first 75 epochs was set to 1 and 100, and after 75 was set to 1000 and 100, respectively.\cite{kovacs2023evaluation,hu2024efficient} \\
For OpenMM and LAMMPS MLFF MD simulation, Nos$\acute{e}$-Hoover thermostat and barostat was used, with relaxation time of 0.1 ps and 0.5 ps for LAMMPS, respectively. 
\subsection{PYR$_{14}$BF$_4$ Ionic Liquid}
While most of the MLFF program packages build a MLFF with a given dataset, DGPEN adopted concurrent learning scheme. After DPMD builds a MLFF, LAMMPS MD simulation (in-built package) is performed and it's performance is measured by comparing the force values of the MD trajectory with the force values calculated from the same structure DFT simulation. If the MLFF fails to predict the force values within the tolerance range, such frames are then added to the training dataset and used in the next training cycle to build an improved MLFF. \\
For the training dataset, 1,040 individual EQ initial configurations and 6,000 individual nEQ initial configurations were prepared, where each configuration composed of 10 ion pairs, density ranging from 0.77 to 1.26 compared to the density obtained from APPLE\&P polarizable FF MD simulation. 40 out of 1,040 EQ structures conducted 100 frame AIMD simulation and rest 1,000 EQ structures conducted single point energy calculation, resulting in a total of 5,000 EQ structure training dataset. 6,000 nEQ structures conducted single point energy calculation which a total of 11,000 EQ\&nEQ structures used in the training dataset. 19,812 structures were added to the training set through out the 7 DPGEN iterations, where a total of 30,812 structures were used to train the final DPGEN MLFF and MACE MLFF.
\subsection{LiTFSI/PYR$_{14}$TFSI Ionic Liquid}
Five types of IL systems were prepared, where the LiTFSI concentration are 0\%, 10\%, 20\%, 30\%, and 40\%, all composed of 10 ion pairs. Each LiTFSI concentration dataset is composed of 1020 individual EQ initial structures and 1,400 individual nEQ initial structures. 20 out of 1,020 EQ structures conducted 100 frame AIMD simulation and rest 1,000 EQ structures conducted single point energy calculation. 400 out of 1,400 nEQ structures conducted 2 frame AIMD simluation and rest 1,000 nEQ structures conducted single point energy calculation. Excluding systems that did not converge or exhibited molecular fragmentation, a total of 22,992 structures were used to train MACE MLFF.

\section{Results and Discussion}
\subsection{PYR$_{14}$BF$_4$ with DPGEN}
\subsubsection{Validation Test}
\begin{figure}[h]
\centering
  \includegraphics[height=11cm]{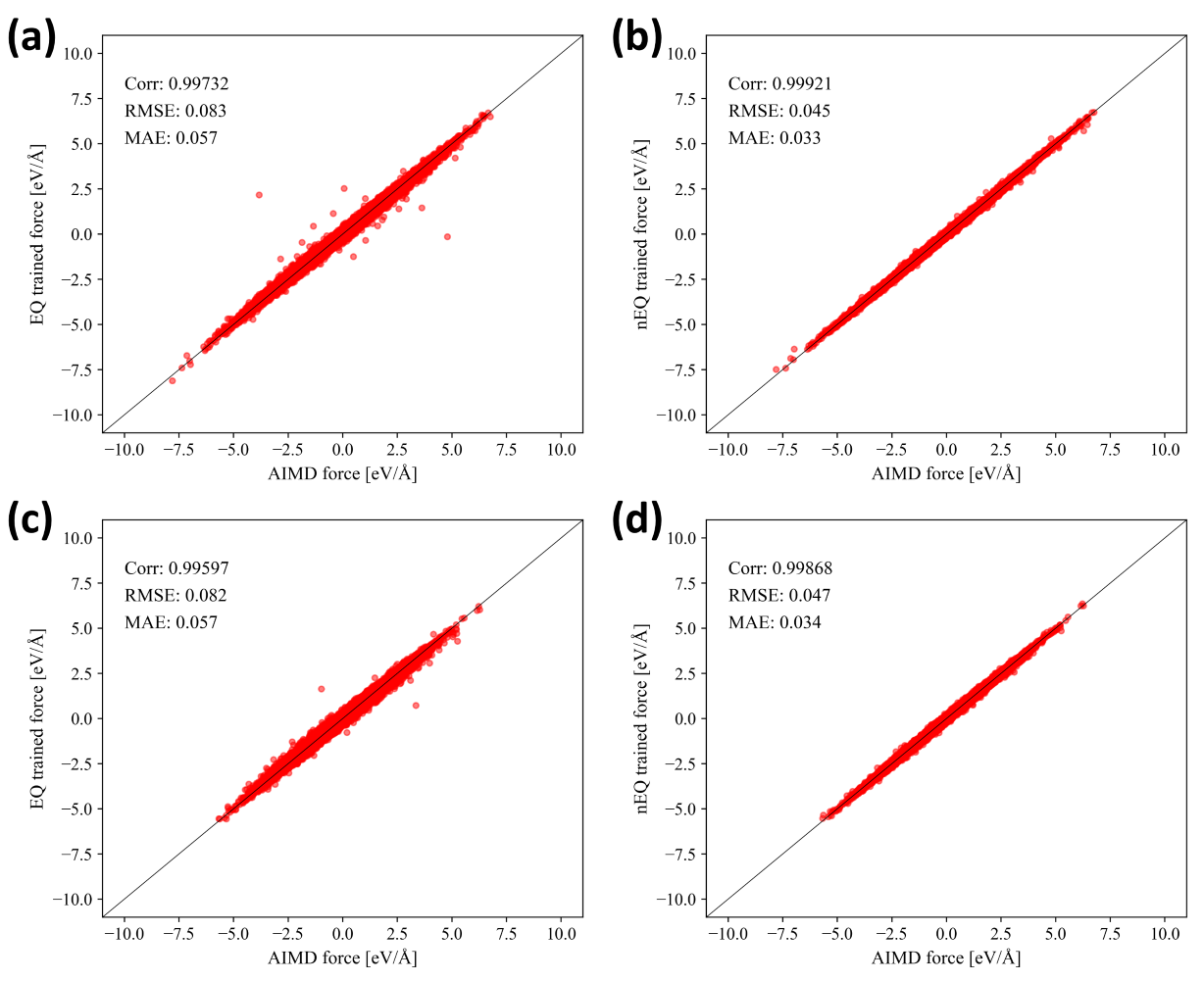}
  \caption{DPMD MLFF verses AIMD force validation tests. EQ validation datasets were tested with (a) EQMLFF (b) nEQMLFF, and nEQ validation datasets were tested with (c) EQMLFF, and (d) nEQMLFF.}
  \label{fgr:EQnEQ_validation}
\end{figure}

Visualization methods may vary depending on the intention of the figure. However, We believe that other than scatter plot, such as mesh grid plots, underestimates the existence of deviating data points that might exist as shown in Figure~\ref{fgr:scatter_bad}. So, in this study, all the energy and force validation plots are shown as scatter plots. \\
The performance of a MLFF depends on how well the training datasets are prepared. Utilizing an long AIMD simulation trajectory, which started from an equilibrium structure, may appear plausible. But the trajectory does not deviates largely from the initial structure as it is basically, a slow AIMD simulation. To test this issue, we prepared two MLFFs and two validation datasets. The two MLFFs were built with DPMD, where one MLFF is built from 40 EQ structures with each 100 frame AIMD simulation performed, a total of 4,000 EQ structures (EQMLFF). The other MLFF is built from single frame energy calculation from 4,000 nEQ structures (nEQMLFF). The two validation datasets are composed of single frame energy calculation of a 200 structure, either EQ or nEQ, and the result is shown in Figure~\ref{fgr:EQnEQ_validation}.

\begin{figure}[h]
\centering
  \includegraphics[height=6cm]{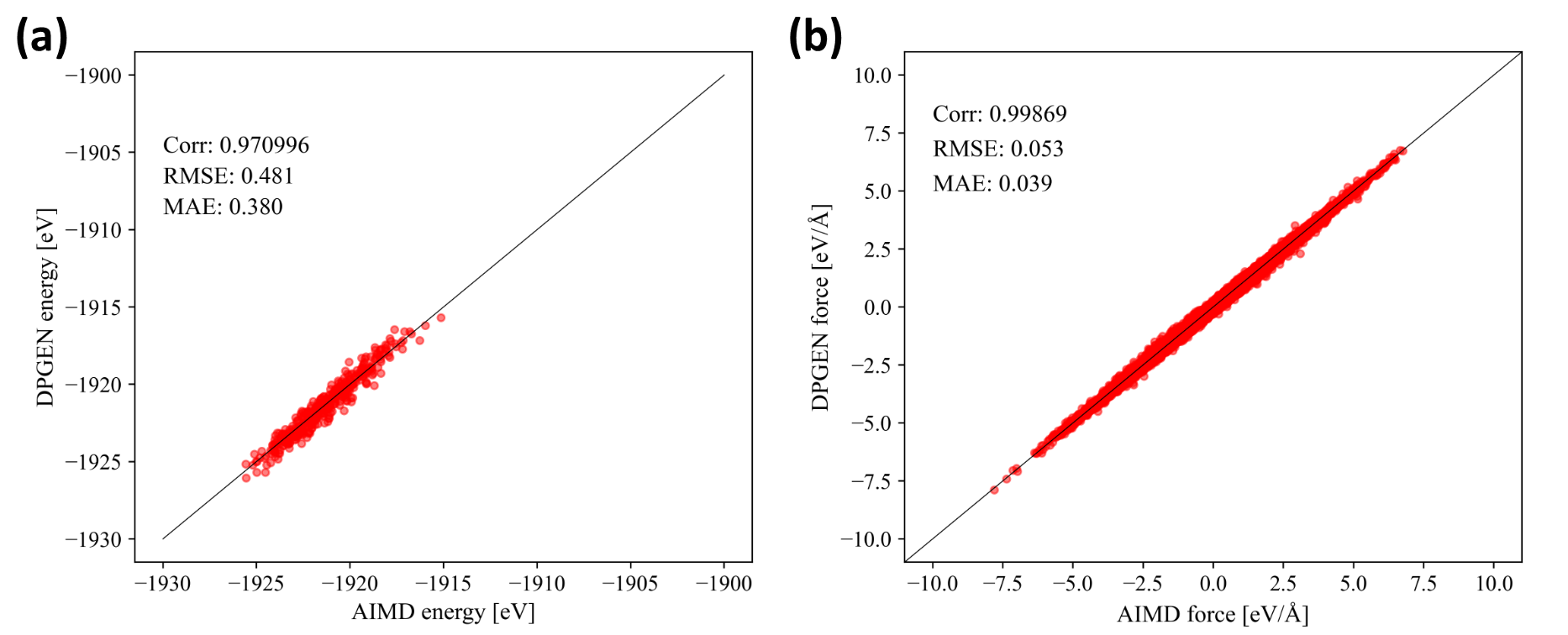}
  \caption{(a) Energy and (b) force validation test (200EQ+200nEQ structure) of final DPMD MLFF.}
  \label{fgr:dpgen_bf4pyr14}
\end{figure}

The performance of nEQMLFF (Figure~\ref{fgr:EQnEQ_validation}(b) and (d)) showed better root-mean-square-error (RMSE) and mean-absolute-error (MAE) than that of EQMLFF (Figure~\ref{fgr:EQnEQ_validation}(a) and (c)). As the purpose of the training dataset is to fill the empty spaces of PES via MLFF, it is nature that that nEQ structures, which structures are largely deviates from each other, was advantageous to efficiently fill those missing PES spaces. But EQ training datasets are essential, as the MD simulation is performed under equilibrium after the system is relaxed. While the bulk system may in a equilibrium state (inter-molecular structure, EQ/nEQ structure), molecule itself may not be in a equilibrium (inter-molecular structure, EQ/nEQ molecule), such as bond length, angle, and dihedral angle.
We gained EQ/nEQ structures simply by performing MD simulation of the system or not. If the initially randomly positioned system underwent very short simulation, it is regarded as nEQ structure. nEQ molecules can be achieved from noising the molecule by imposing constraints to a molecule.~\cite{ni2024pre} But in this study, we simply obtained nEQ molecules from high temperature and large time step (dt). We tuned the temperature and time step just before the molecule decomposition happens, but high enough to scan over various bond lengths and angles. Example of such AIMD simulation is listed in Supplementary video file. The validation test results of each DPGEN MLFF iteration is shown in Figure~\ref{fgr:validEQ_E}~\ref{fgr:validnEQ_F}, and the final validation tests of DPMD trained from 30,811 structures are shown in Figure~\ref{fgr:dpgen_bf4pyr14}, which we obtained force deviation MAE and RMSE of 0.039 eV/Å and 0.053 eV/Å, respectively.

\subsubsection{Molecular Dynamics Simulation}
If a MLFF is improperly constructed, several prominent errors can be immediately identified, such as formation of vacuum under NPT simulation, unintended decomposition of molecule, or the system just collapse and form a particle-aggregated ball which looks like high-entropy alloy nanoparticle. We should note that such incidents were not observed during DPMD MLFF MD simulation. First, dynamics of the system was evaluated through diffusion coefficients calculated from mean square distance (MSD) (Figure~\ref{fgr:DPGEN_MSD}).
\begin{equation}
  D=\frac{1}{6N}\lim_{t\rightarrow\infty}\frac{d}{dt}\sum^N_{i=1}\langle|r_i(t)-r_i(0)|^2\rangle
  \label{eq:D_MSD}
\end{equation}
Diffusion coefficients of BF$_4$PYR$_{14}$ IL is calculated using Equation~\ref{eq:D_MSD}, where $N, r(t), \langle\rangle$ refers to the number of atoms, position vector of the atom at time t, and the ensemble average, respectively. Calculated diffusion coefficients are summarized in Table~\ref{tbl:Diffusion1}. DPMD MLFF diffusion coefficients for both anion and cation ($\sim0.6$) was faster than the values calculated from CL\&P ($~0.1$), but slower than APPLE\&P ($\sim2.3$). IL density calculated from each FFs is also listed in Table~\ref{tbl:Density1}. While the density values of CL\&P (1.03) and APPLE\&P (1.07) are similar, density value obtained from DPMD (1.28) was exceptionally high. It is interesting that the dynamics from DPMD MD simulation was faster than CL\&P MD simulation though the density is much higher than that of CL\&P.
\begin{table}
\centering
  \caption{Diffusion coefficients of PYR$_{14}$ and BF$_4$ calculated from four force fields.}
  \label{tbl:Diffusion1}
  \begin{tabular}{cccc}
    \hline
    & CL\&P & APPLE\&P & DPGEN \\
    \hline
    \multirow{2}{*}{BF$_4$}&0.1445&2.2999&0.5859 \\
    &$\pm$0.0008&$\pm$0.0180&$\pm$0.0024\\
    \hline
    \multirow{2}{*}{PYR$_{14}$}&0.1280&1.8594&0.5851\\
    &$\pm$0.0001&$\pm$0.0081&$\pm$0.0007\\
    \hline
    \multicolumn{4}{l}{*Unit x10$^{-10}m^2s^{-1}$}\\
  \end{tabular}
\end{table}
\begin{table}
\centering
  \caption{Density of BF$_4$PYR$_{14}$ IL at 423K calculated from following force fields.}
  \label{tbl:Density1}
  \begin{tabular}{cc}
    \hline
    & Density \\
    \hline
    CL\&P&1.0345$\pm$0.0003\\
    APPLE\&P&1.0707$\pm$0.0006\\
    DPMD&1.2800$\pm$0.0039\\
    \hline
    \multicolumn{2}{l}{*Unit $gcm^{-3}$}\\
  \end{tabular}
\end{table}
Further structural analysis was performed using radial distribution function (RDF) to compare the structural differences between the FFs. 
\begin{equation}
  g(r)=\frac{1}{\langle\rho\rangle_{local}}\sum\frac{\delta(r_{ij}-r)}{4\pi r^2}
  \label{eq:RDF}
\end{equation}
RDF is calculated from Equation~\ref{eq:RDF}, where $\langle\rho\rangle, r_{ij}, \delta(r_{ij}-r)$ refers to the averaged atom density, the distance between atom $i$ and $j$, and the Dirac delta function which equals one when $r_{ij}$ equals $r$. RDFs are shown in Figure~\ref{fgr:BF4PYR14_RDF} and differences among the FFs could be observed, such as only B(BF$_4$)-B(BF$_4$) DPMD RDF (Figure~\ref{fgr:BF4PYR14_RDF}(c)) showed a shoulder peak at 6 Å. However, due to DPMD's exceptionally high density and low dynamic properties, we determined that the system was fundamentally flawed from the outset. Therefore, no further analysis was conducted. 

\subsection{Ionic Liquid with MACE}
BF$_4$PYR$_{14}$ and LiTFSI/PYR$_{14}$TFSI IL MLFF were built using MACE, where same training dataset when building PYR$_{14}$BF$_4$ MLFF using DPMD was used, and the training dataset for LiTFSI/PYR$_{14}$TFSI is listed in Method section.
\subsubsection{Validation Test}
\begin{figure}[h]
\centering
  \includegraphics[height=6cm]{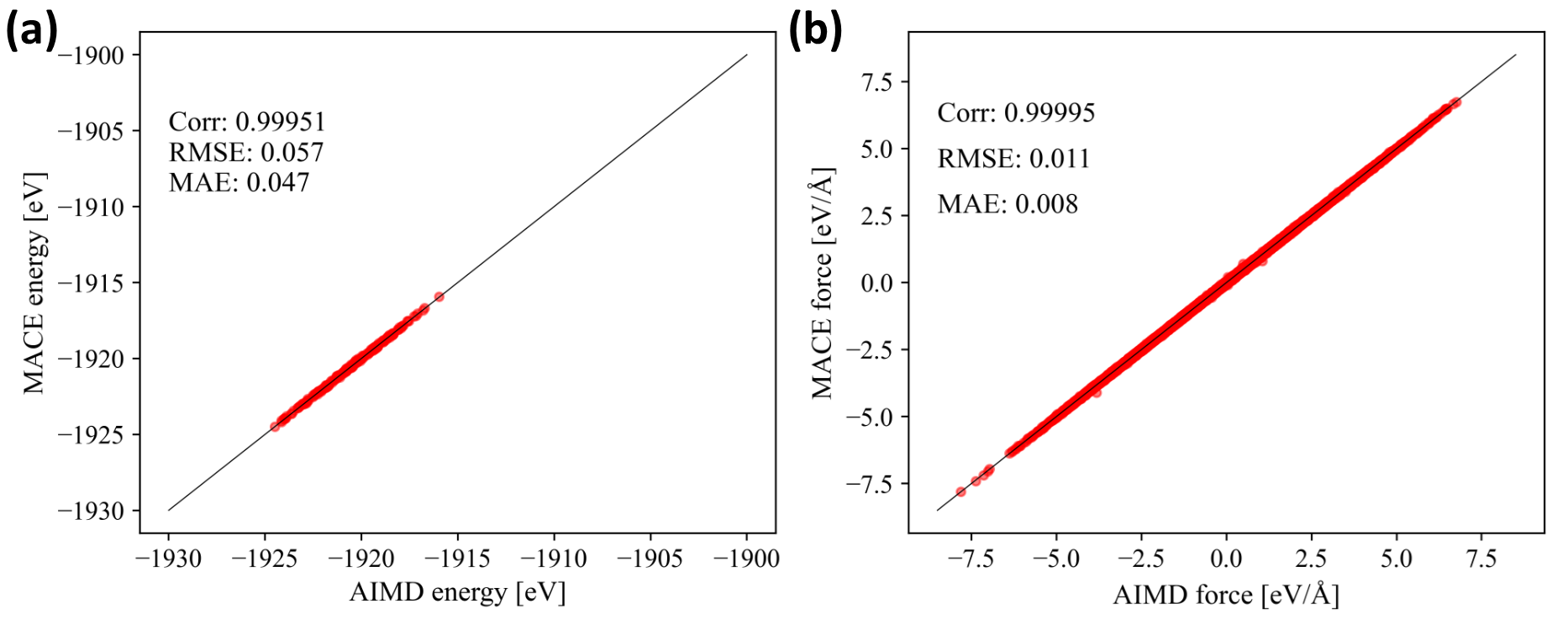}
  \caption{(a) Energy and (b) force validation test of MACE MLFF.}
  \label{fgr:mace_bf4}
\end{figure}

\begin{figure}[h]
\centering
  \includegraphics[height=6cm]{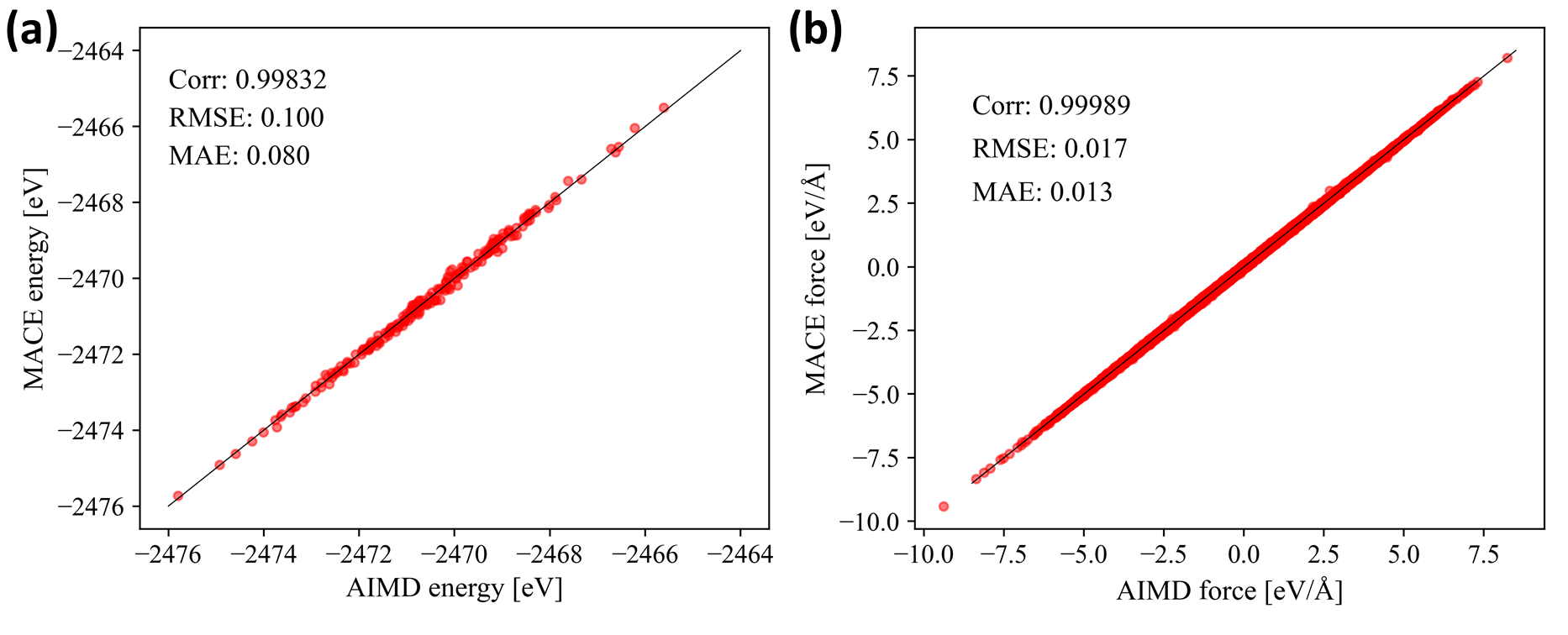}
  \caption{(a) Energy and (b) force validation test of MACE MLFF with TFSIPYR$_{14}$ binary IL.}
  \label{fgr:trainLi0}
\end{figure}

Same validation test set which was tested to DPMD was used with MACE MLFF validation, and the results are shown in Figure~\ref{fgr:mace_bf4}. We achieved high force accuracy, where RMSE and MAE of 0.011 eV/Å and 0.008 eV/Å, respectively. It is also interesting that though MACE model (GNN) is fundamentally different from DPMD model (DNN), energy prediction accuracy could be significantly increased through two stage weight method as shown in previous studies.\cite{kovacs2023evaluation,hu2024efficient}
MLFF for LiTFSI/PYR$_{14}$TFSI IL was built MACE using 22,992 structures (each composed of 10 ion pairs) and the validation test results of LiTFSI concentration 0\% (binary IL), 10\%, 20\%, 30\%, and 40\% are shown in Figure~\ref{fgr:trainLi0} and Figure~\ref{fgr:trainLi1}$\sim$~\ref{fgr:trainLi4}.
As the training dataset is generally evenly distributed among the LiTFSI concentration 0\%$\sim$40\%, We expected low accuracy on the validation test set of LiTFSI concentration 0\% and 40\%. Nevertheless, contrary to initial expectations, both exhibited excellent prediction accuracy as shown in Figure~\ref{fgr:trainLi0} and Figure~\ref{fgr:trainLi4}. Note that each validation test set is composed of sum of 200 individual EQ structures and 200 individual nEQ structures, total of 400 structures with density varies from -20\% to +20\%, which are totally irrelevant from training datasets.
\subsubsection{Molecular Dynamics Simulation}
\begin{figure}[h]
\centering
  \includegraphics[height=6cm]{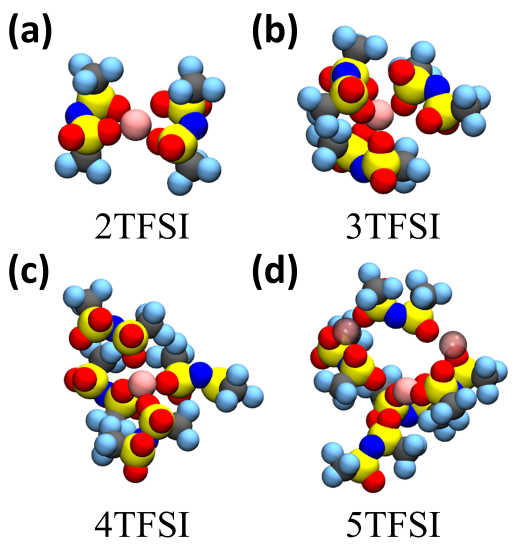}
  \caption{Li cluster that coordinating with (a) 2 TFSIs, (b) 3 TFSIs, (c) 4 TFSIs, and (d) 5 TFSIs within the first shell 6 \AA. 5$^{th}$ TFSI in 5TFSI does not directly coordinate with Li, but detected within the first shell.}
  \label{fgr:cluster}
\end{figure}

The dynamic properties of MACE MLFF was also compared with CL\&P, APPLE\&P and the results are summarized in Table~\ref{tbl:Dmace}. Diffusion coefficients increased with the FF order CL\&P, APPLE\&P, and MACE MLFF. It is interesting that with same training dataset, diffusion coefficient of anion/cation drastically increased from 0.586/0.585 to 3.166/2.616 x$10^{-10}m^2s^{-1}$ with the change of MLFF building package from DPMD to MACE.
\begin{sidewaystable}
\centering
  \caption{Diffusion coefficients of PYR$_{14}$BF$_4$ and LiTFSI/PYR$_{14}$TFSI ILs calculated from CL\&P, APPLE\&P, and MACE MLFF.}
  \label{tbl:Dmace}
  \begin{tabular}{ccccccccccccccc}
    \cline{2-13}
    \multirow{2}{*}{System}&\multicolumn{3}{c}{CL\&P}&\multicolumn{3}{c}{APPLE\&P}&\multicolumn{3}{c}{MACE}&\multicolumn{3}{c}{Exp}\\
    ani&cat&Li&ani&cat&Li&ani&cat&Li&ani&cat&Li\\
    \hline
    \multirow{2}{*}{PYR$_{14}$BF$_4$}&\multirow{2}{*}{0.145}&\multirow{2}{*}{0.128}&\multirow{2}{*}{-}&\multirow{2}{*}{2.300}&\multirow{2}{*}{1.859}&\multirow{2}{*}{-}&\multirow{2}{*}{3.166}&\multirow{2}{*}{2.616}&\multirow{2}{*}{-}&\multirow{2}{*}{-}&\multirow{2}{*}{-}&\multirow{2}{*}{-}\\
    \\
    \cline{2-13}
    \multirow{2}{*}{TFSIPYR$_{14}$}&\multirow{2}{*}{0.322}&\multirow{2}{*}{0.296}&\multirow{2}{*}{-}&\multirow{2}{*}{2.498}&\multirow{2}{*}{2.739}&\multirow{2}{*}{-}&\multirow{2}{*}{2.771}&\multirow{2}{*}{3.537}&\multirow{2}{*}{-}&\multirow{2}{*}{4.710$^a$}&\multirow{2}{*}{5.136$^a$}&\multirow{2}{*}{-}\\
    \\
    \cline{2-13}
    \multirow{2}{*}{0.1LiTFSI}& \\
    \multirow{2}{*}{423K}&\multirow{2}{*}{0.201}&\multirow{2}{*}{0.245}&\multirow{2}{*}{0.087}&\multirow{2}{*}{2.026}&\multirow{2}{*}{2.421}&\multirow{2}{*}{1.273}&\multirow{2}{*}{2.419}&\multirow{2}{*}{3.553}&\multirow{2}{*}{1.458}&\multirow{2}{*}{3.163$^b$}&\multirow{2}{*}{3.207$^b$}&\multirow{2}{*}{2.958$^b$}\\
    \multirow{2}{*}{353K}&\multirow{2}{*}{0.016}&\multirow{2}{*}{0.018}&\multirow{2}{*}{0.008}&\multirow{2}{*}{0.453}&\multirow{2}{*}{0.597}&\multirow{2}{*}{0.284}&\multirow{2}{*}{0.693}&\multirow{2}{*}{0.992}&\multirow{2}{*}{0.475}&\multirow{2}{*}{0.660$^b$}&\multirow{2}{*}{0.815$^b$}&\multirow{2}{*}{0.521$^b$}\\
    \multirow{2}{*}{323K}&\multirow{2}{*}{0.007}&\multirow{2}{*}{0.008}&\multirow{2}{*}{0.004}&\multirow{2}{*}{0.191}&\multirow{2}{*}{0.226}&\multirow{2}{*}{0.105}&\multirow{2}{*}{0.493}&\multirow{2}{*}{0.596}&\multirow{2}{*}{0.207}&\multirow{2}{*}{0.235$^b$}&\multirow{2}{*}{0.316$^b$}&\multirow{2}{*}{0.165$^b$}\\
    \\
    \cline{2-13}
    \multirow{2}{*}{0.2LiTFSI}&\multirow{2}{*}{0.076}&\multirow{2}{*}{0.111}&\multirow{2}{*}{0.034}&\multirow{2}{*}{1.574}&\multirow{2}{*}{2.428}&\multirow{2}{*}{1.603}&\multirow{2}{*}{}&\multirow{2}{*}{}&\multirow{2}{*}{}&\multirow{2}{*}{-}&\multirow{2}{*}{-}&\multirow{2}{*}{-}\\
    \\
    \cline{2-13}
    \multirow{2}{*}{0.3LiTFSI}&\multirow{2}{*}{0.048}&\multirow{2}{*}{0.065}&\multirow{2}{*}{0.023}&\multirow{2}{*}{1.218}&\multirow{2}{*}{1.701}&\multirow{2}{*}{1.013}&\multirow{2}{*}{1.481}&\multirow{2}{*}{2.805}&\multirow{2}{*}{1.023}&\multirow{2}{*}{-}&\multirow{2}{*}{-}&\multirow{2}{*}{-}\\
    \\
    \hline
    \multicolumn{7}{l}{*Unit x10$^{-10}m^2s^{-1}$}\\
    \multicolumn{7}{l}{*VFT diffusion coefficient from a: ref~\cite{tokuda2006physicochemical} and b: ref~\cite{solano2013joint}}\\
  \end{tabular}
\end{sidewaystable}
System density through FFs were also calculated and listed in Table~\ref{tbl:Density2}. While MACE MLFF reasonably predicted the density of PYR$_{14}$BF$_4$, and PYR$_{14}$TFSI within the error range 1\% compared to experimental results, MACE MLFF seemed to underestimated the IL density when Li ion was introduced. 
\begin{table}
\centering
  \caption{Density of BF$_4$PYR$_{14}$ IL at 423K calculated from following force fields.}
  \label{tbl:Density2}
  \begin{tabular}{ccccc}
    \hline
    System&CL\&P&APPLE\&P&MACE&Exp\\
    \hline
    PYR$_{14}$BF$_4$&1.0345&1.0707&1.0443&-\\
    PYR$_{14}$TFSI&1.2795&1.3035&1.2722&1.2855$^a$\\
    0.1LiTFSI\\
    423K&1.3075&1.3290&1.2778&1.3091$^b$\\
    353K&1.3693&1.3853&1.3224&1.3717$^b$\\
    323K&1.3952&1.4111&1.3390&1.3986$^b$\\
    0.2LiTFSI&1.3412&1.3578&1.2983&-\\
    0.3LiTFSI&1.3791&1.3896&1.3300&-\\
    \hline
    \multicolumn{5}{l}{*Unit $gcm^{-3}$}\\
    \multicolumn{5}{l}{*VFT density from a: ref~\cite{tokuda2006physicochemical} and b: ref~\cite{solano2013joint}}\\
  \end{tabular}
\end{table}
To further investigate the local structures in LiTFSI/PYR$_{14}$TFSI IL system,Li cluster population analysis was performed, where Li clusters are categorized by the number of anions (TFSI) coordinating within the first shell. Majority of Li cluster (>99\%) favorably interact with 4 oxygen originated from TFSIs, forming a tetrahedron structure. So the number of TFSI that can coordinate with Li range from 2 to 4. The example Li cluster image is shown in Figure~\ref{fgr:cluster} and their detailed population are summarized in Table~\ref{tbl:Pop1}.
\begin{table}
\centering
  \caption{The Li cluster population distribution from CL\&P, APPLE\&P, and MACE MLFF MD simulation categorized based on the number of TFSI molecules coordinated to Li within the first shell 6 \AA.}
  \label{tbl:Pop1}
  \begin{tabular}{cccccccccc}
    \hline
    Number of&\multicolumn{3}{c}{CL\&P}&\multicolumn{3}{c}{APPLE\&P}&\multicolumn{3}{c}{MACE}\\
    TFSI&2&3&4&2&3&4&2&3&4\\
    \hline
    0.1Li423K&0.00&79.19&20.52&6.31&85.22&8.44&12.40&86.44&1.17\\
    0.1Li353K&0.00&86.61&13.37&2.51&92.30&5.20&11.37&86.43&2.21\\
    0.1Li323K&0.00&80.71&19.08&1.76&91.80&6.43&6.41&88.89&4.71\\
    0.2Li&0.00&62.38&36.71&5.16&79.49&15.14&11.09&85.80&3.10\\
    0.3Li&0.00&45.97&51.69&5.94&77.75&16.02&12.56&81.11&6.30\\
    \hline
  \end{tabular}
\end{table}
CL\&P showed the biggest difference among three FFs that it had no 2TFSI Li cluster among all the simulated ILs. Li cluster populations were similar with APPLE\&P and MACE, while MACE predicted more 2TFSI cluster than APPLE\&P and APPLE\&P predicted more 4TFSI cluster than MACE.

\subsection{Molecule Structure Analysis}
\begin{figure}[h]
\centering
  \includegraphics[height=6cm]{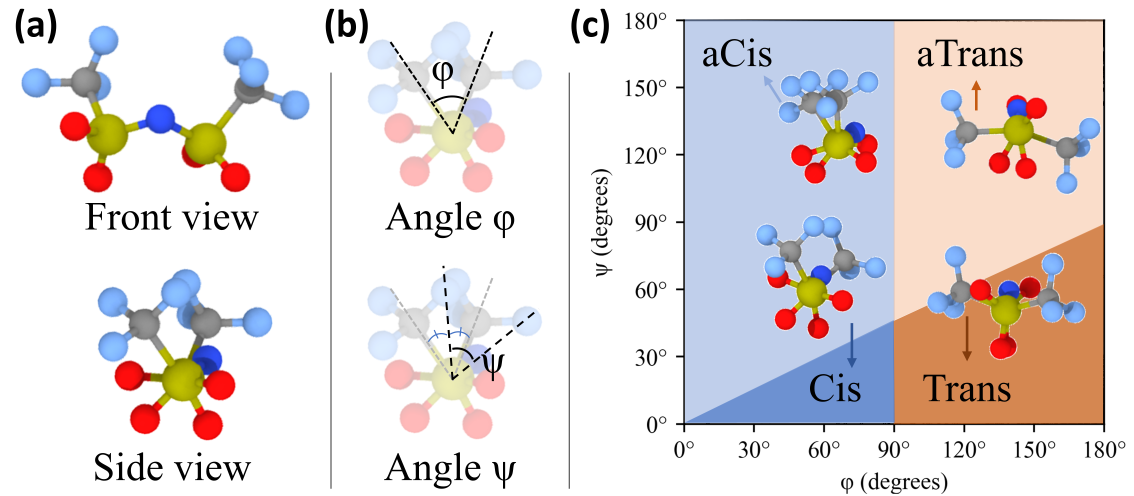}
  \caption{Image of (a) TFSI, (b) justification of TFSI shape by angle $\phi$ and $\psi$, and (c) example TFSI structure of $Cis$, $aCis$, $Trans$, and $aTrans$ classified from $\phi$ and $\psi$.}
  \label{fgr:tfsi_explain}
\end{figure}

Shape of TFSI and PYR$_{14}$, each can be determined via two variables. The classification of TFSI shape is widely studied and known as cis and trans shape determined from C-S-S-C dihedral angle as shown in Figure~\ref{fgr:tfsi_explain}(b) $\phi$.\cite{li2008molecular,holbrey2004crystal} In this study, we have taken the position of nitrogen (N) into account, so that the exact TFSI configuration can be determined from two variables. This way, N can position either in-between the dihedral angle or outer side of the dihedral angle, which we named it as $Cis$, $aCis$, $Trans$, and $aTrans$ as shown in Figure~\ref{fgr:tfsi_explain}(c). Detailed description of the TFSI structure corresponding to the points on the graph is shown in Figure~\ref{fgr:tfsi_detail}. Note that the transition between $Trans$ and $aTrans$ occur from just a small undulation, flapping of the molecule. We should note that our trained MLFF, MPA-0, and Omat-0 all runs PYR$_{14}$TFSI md simulation without any eruption during the simulation run. Also the training dataset of our trained MLFF only includes bulk systems, which has zero information of how a single molecule would act in a vacuum state.
\begin{figure}[h]
\centering
  \includegraphics[height=8cm]{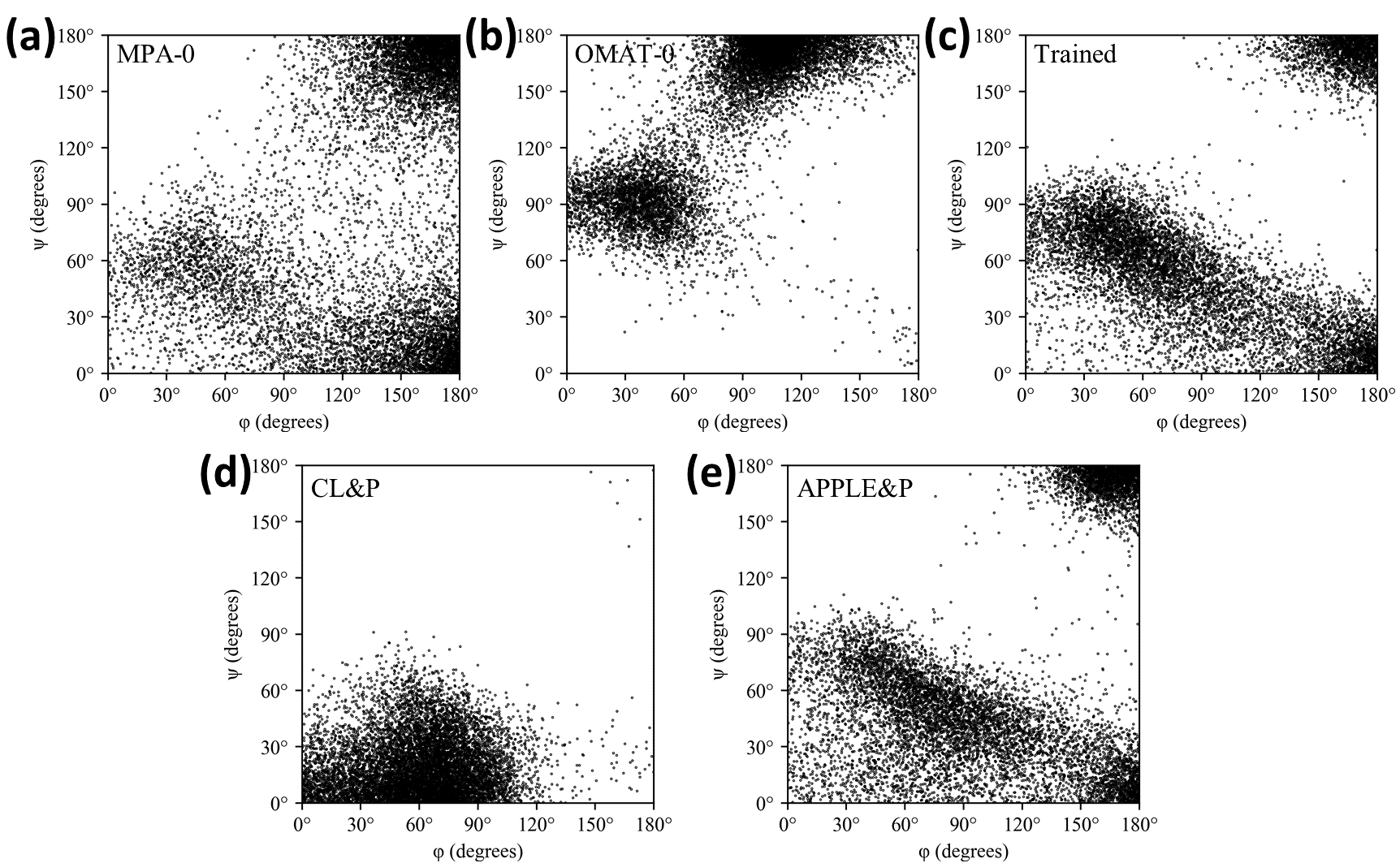}
  \caption{Dot distribution plots of a single TFSI in a vacuum environment simulated with (a) MPA-0, (b) OMAT-0, (c) trained MLFF, (d) CL\&P, and (e) APPLE\&P force field.}
  \label{fgr:tfsi_shape}
\end{figure}
Dot distribution plots of a single TFSI configuration in vacuum state under various force fields are shown in Figure~\ref{fgr:tfsi_shape}. It is noteworthy that the structure distribution from APPLE\&P was almost same to that of our trained MLFF (MACE). At both force fields, we can observe smooth transitions from $aCis$, $Trans$ to $aTrans$. Distribution of MPA-0 is somewhat similar to APPLE\&P, but it underestimate the $aCis$ and has a tendency to make a broad or generalized predictions. OMAT-0 shows a different distribution that exhibited almost no $Cis$ and $Trans$ structures, rather TFSI transitioned directly from $aCis$ to $aTrans$. For this phenomenon to occur, the central nitrogen atom of TFSI should exist in a twisted state and the C-S-S-C dihedral angle should rotate in the direction opposite to that of the previously rotating dihedral angle. CL\&P showed strong preference of $Cis$ structure with some $Trans$ structures originated from molecule fluctuation. Such results imply that interpretation of structural properties and dynamics from CL\&P and OMAT-0 can be totally wrong, in case the APPLE\&P and our trained MLFF results are correct.

\begin{figure}[h]
\centering
  \includegraphics[height=9cm]{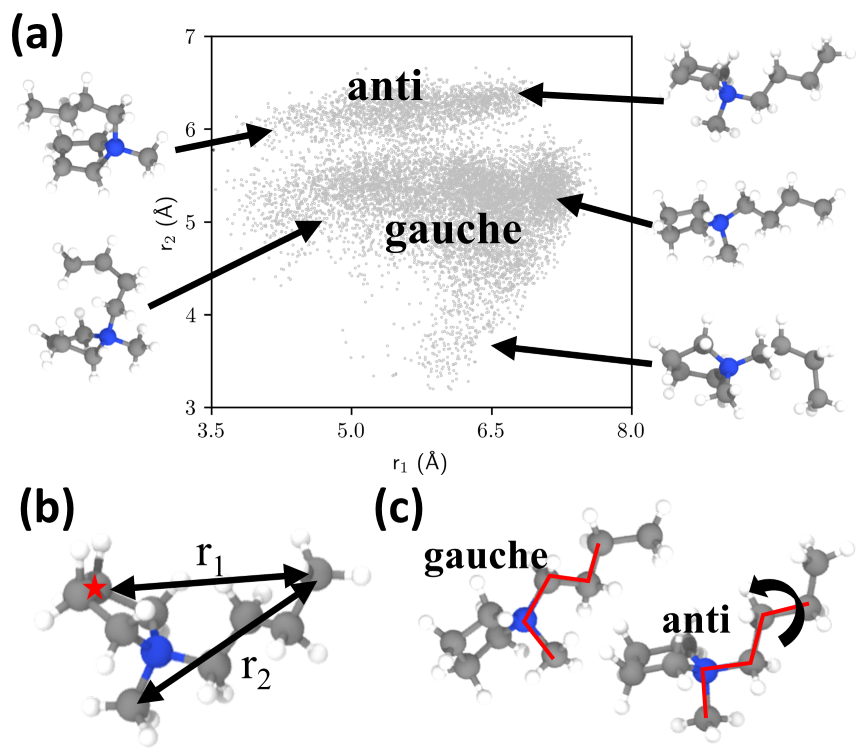}
  \caption{Image of (a) TFSI, (b) justification of TFSI shape by angle $\phi$ and $\psi$, and (c) example TFSI structure of $Cis$, $aCis$, $Trans$, and $aTrans$ classified from $\phi$ and $\psi$.}
  \label{fgr:pyr_explain}
\end{figure}

In case of PYR$_{14}$, we choose the end-to-end distance of butane tail with methyl ($r_2$) and carbon atom from the pentagon ring ($r_1$), as shown in Figure~\ref{fgr:pyr_explain}(b). Example structures of PYR$_{14}$ with varying $r_1$ and $r_2$ are shown in Figure~\ref{fgr:pyr_explain}(a). 
Note that a clear transition is observed near $r_2=6$Å, where the butane tail structure changes from $anti$ to $gauche$ as illustrated in Figure~\ref{fgr:pyr_explain}(c). 
\begin{figure}[h]
\centering
  \includegraphics[height=8cm]{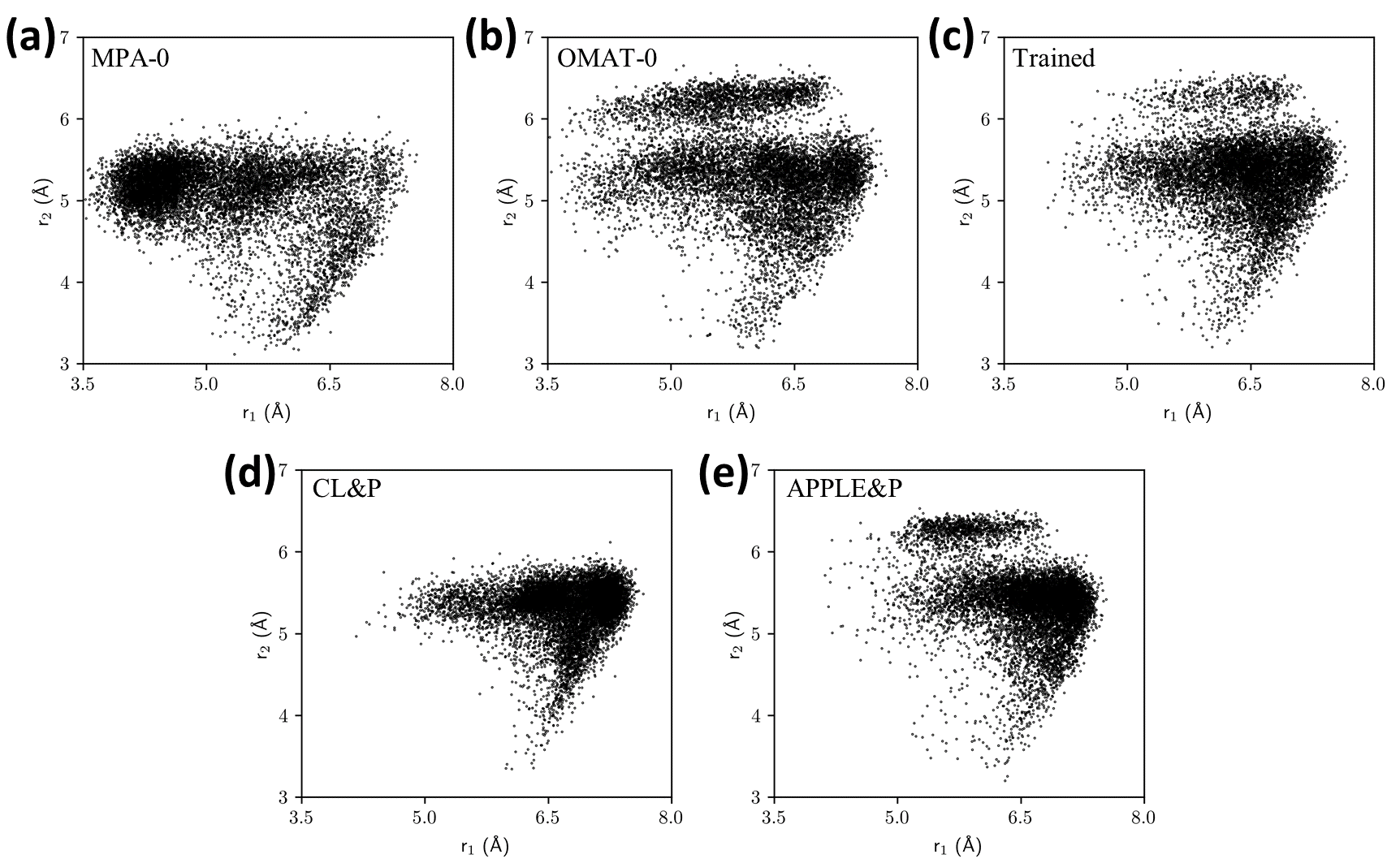}
  \caption{Dot distribution plots of a single PYR$_{14}$ in a vacuum environment simulated with (a) MPA-0, (b) OMAT-0, (c) trained MLFF, (d) CL\&P, (e) APPLE\&P force field.}
  \label{fgr:pyr_shape}
\end{figure}
PYR$_{14}$ structure distribution along varying force fields are shown in Figure~\ref{fgr:pyr_shape} and PES calculated from DFT simulation is shown in Figure~\ref{fgr:PYR_PES}. In case of CL\&P, where molecular structure is controlled by strong angle and dihedral potential, PYR$_{14}$ butane tail could not overcome the potential barrier to transit from $gauche$ structure to $anti$ structure so that the whole distribution is concentrated in $gauche$ structure region. MPA-0 also shows similar $gauche$ preference behavior, but showed a distribution concentrated at the position $r_1=4$Å. Unique feature of such structure is that butane tail is bended toward pyrrolidine ring structure like a shape of scorpion. While such self-interacting structure is prevented in CL\&P and APPLE\&P by structural hindrance, MPA-0 missed such hindrances originated from complex molecular structure. OMAT-0 and our trained MLFF both captured all $gauche$ and $anti$ structure, while OMAT-0 overestimated self-interacting scorpion-like structure ($r_1\sim4$Å) and our trained MLFF underestimated the stability of $anti$ structure compared to the results from APPLE\&P.
From the fact that the molecular shape that our trained MLFF described surprisingly resembled with that obtained from APPLE\&P force field, which is known to be the most accurate among the force fields developed to date, we believe that the molecular distribution that our trained MLFF and APPLE\&P showed would resemble with the true answer. Under this assumption, MPA-0 roughly succeeded in predicting the TFSI structure but failed to predict PYR$_{14}$ structure, whereas OMAT-0 exhibited the opposite trend, failing to predict the TFSI structure while successfully predicting the PYR$_{14}$ structure. Meanwhile, CL\&P both failed to predict the structure of TFSI and PYR$_{14}$

\section{Conclusions}
\begin{figure}[h]
\centering
  \includegraphics[height=8cm]{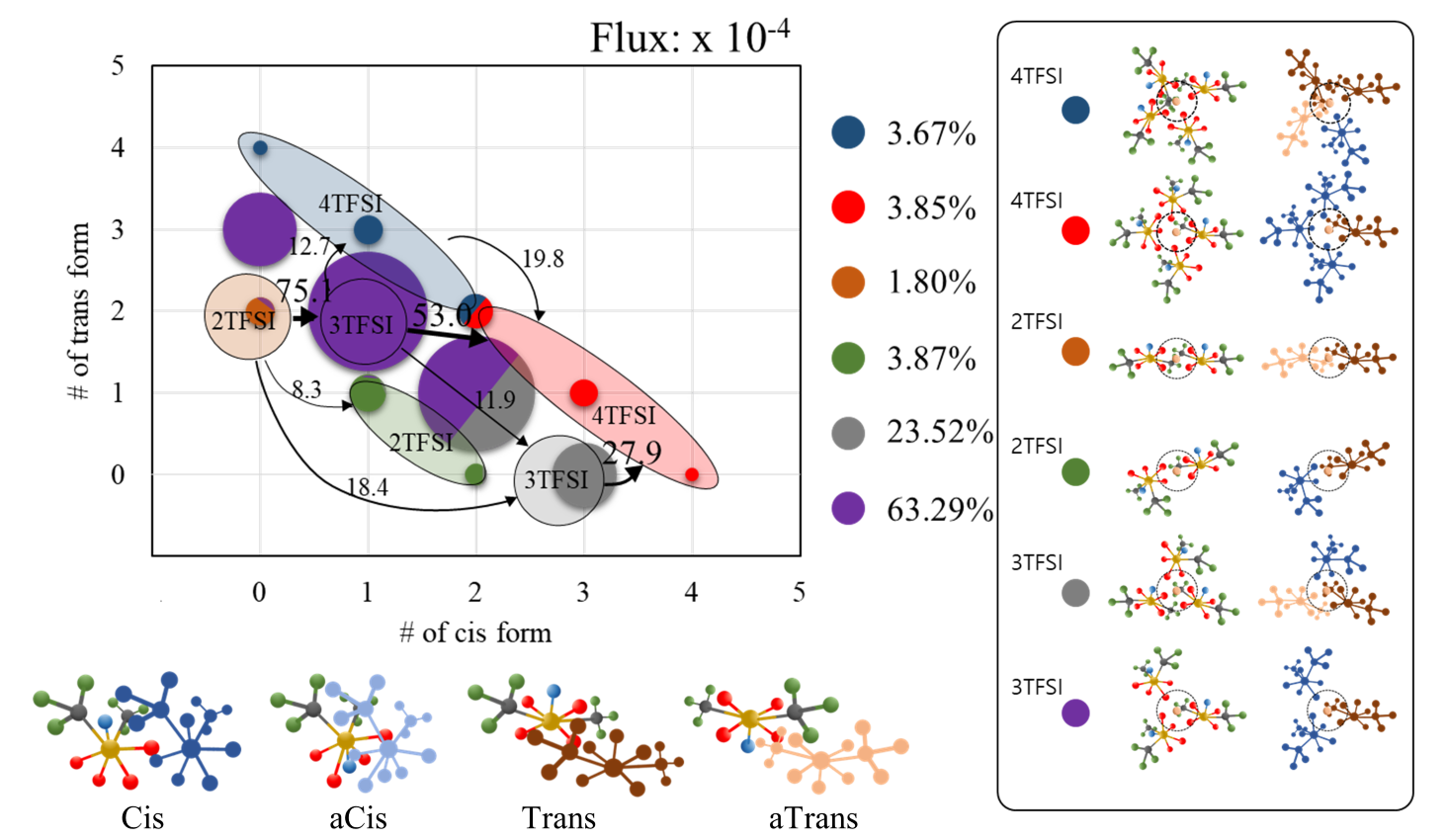}
  \caption{Example of Li cluster transition using Markov State Modeling, where Li cluster is categorized by the number of TFSI and it's shape.}
  \label{fgr:msm}
\end{figure}

In this study, we discussed importance of detailed preparation of the MLFF training dataset and built MLFF was evaluated by structural and dynamical properties. While fine-tuning hyper parameters of a MLFF building program will increase the accuracy of the built MLFF, we failed to achieve a better validation performance to a certain point with IL PYR$_{14}$BF$_4$ with DPMD. DPMD MLFF overestimated the attraction force between the molecules so that the density was 20\% more higher than that from APPLE\&P force field, resulted in a slow dynamics of the overall system. However with the same training datasets, MACE MLFF successfully predicted the density and dynamics similar to the values from APPLE\&P force field. MACE MLFF also successfully predicted the properties of binary IL PYR$_{14}$TFSI, but underestimated the density of the system when Li was introduced. Also, Structural property was, Such claim is supported by the Li cluster analysis, that while our trained MACE MLFF and APPLE\&P force field both mostly preferred Li clusters composed of 3 TFSIs, second most preferred cluster was those composed of 4 TFSIs in case of APPLE\&P and 2 TFSIs in case of our trained MLFF, resulting in sparser clusters and density.
From single molecule simulation analysis, we found that the structure diversity that TFSI and PYR$_{14}$ could have with our trained MLFF was surprisingly similar to that obtained from APPLE\&P force field, indicating that the superiority of APPLE\&P not only achieve dynamical properties of IL, but also the structural details at an DFT level accuracy.~\cite{bedrov2019molecular} In case of MPA-0 and OMAT-0 without fine-tuning, however, each failed to replicate the $anti$ PYR$_{14}$ structure and $Trans$ TFSI structure, respectively.
Further studies should be done with Li cluster lifetime and transitions to deeply understand how MLFF modifies the structure and dynamics in the MD simulation. By classifying the Li clusters by the number of TFSI coordinating with Li, TFSI shape and the number of oxygen that coordinating with Li, we can identify the whole Li transition map as shown in Figure~\ref{fgr:msm}

\section{Conflicts of interest}
There are no conflicts to declare.

\begin{acknowledgement}

\end{acknowledgement}

\begin{suppinfo}
The following files are available free of charge.
\begin{itemize}
  \item suppInfo.pdf: Supporting Figures
\end{itemize}

\end{suppinfo}

\bibliography{refs}

\end{document}


\pagestyle{fancy}
\thispagestyle{plain}
\fancypagestyle{plain}{
\renewcommand{\headrulewidth}{0pt}}

\makeatletter 
\newlength{\figrulesep} 
\setlength{\figrulesep}{0.5\textfloatsep} 
\newcommand{\topfigrule}{\vspace*{-1pt}%
\noindent{\color{cream}\rule[-\figrulesep]{\columnwidth}{1.5pt}} }
\newcommand{\botfigrule}{\vspace*{-2pt}%
\noindent{\color{cream}\rule[\figrulesep]{\columnwidth}{1.5pt}} }
\newcommand{\dblfigrule}{\vspace*{-1pt}%
\noindent{\color{cream}\rule[-\figrulesep]{\textwidth}{1.5pt}} }
\begin{titlepage}
\begin{center}
\vspace*{1cm}
\LARGE
\textbf{Supporting Information:\\Ionic Liquid Molecular Dynamics Simulation with Machine Learning Force Fields: DPMD and MACE}\\
\vspace{3cm}
\large
\textbf{ Anseong Park,\textit{$^{a}$}, Jaeyune Ryu,\textit{$^{a}$} Won Bo Lee,$^{\ast}$\textit{$^{a}$}} \\
\vspace{1cm}
\normalsize
$^{a}$School of Chemical and Biological Engineering and Institute of Chemical Processes, Seoul National University, Seoul 08826, Republic of Korea. Tel: +82 (0)2 880 1529; E-mail: wblee@snu.ac.kr\\
\end{center}
\end{titlepage}

\begin{figure}[p]
\centering
  \includegraphics[height=6cm]{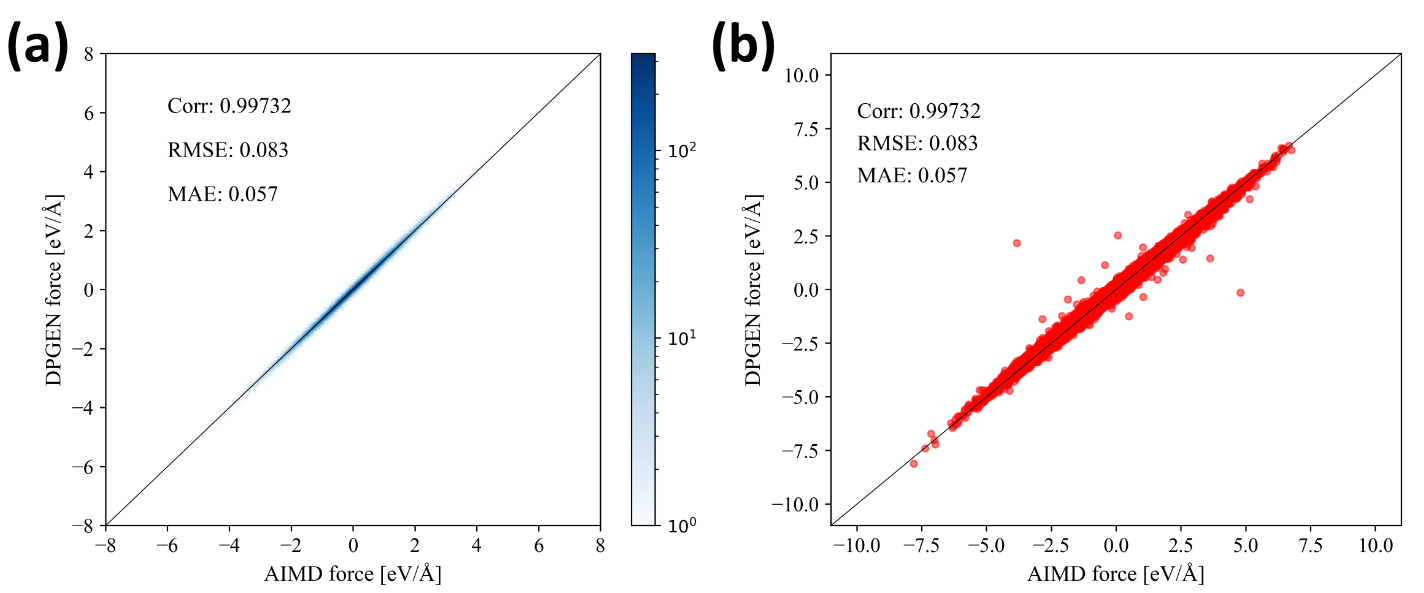}
  \caption{MLFF and AIMD force validation plots drawn by (a) mesh drawing method and (b) scatter method.}
  \label{fgr:scatter_bad}
\end{figure}

\begin{table}[p]
\centering
  \caption{Summary of the overall MD simulation.}
  \label{tbl:sys_summary}
  \begin{tabular}{ccccc}
    \hline
    Tool & Force Field & Component & T (K) & Run (ns) \\
    \hline
    \multirow{7}{*}{GROMACS} & \multirow{7}{*}{CL\&P} & 300PYR$_{14}$BF$_4$ & 423 & 140\\
    &&300PYR$_{14}$TFSI & 423 & 70\\
    &&30LiTFSI/270PYR$_{14}$TFSI & 423 & 70\\
    &&30LiTFSI/270PYR$_{14}$TFSI & 353 & 140\\
    &&30LiTFSI/270PYR$_{14}$TFSI & 323 & 140\\
    &&60LiTFSI/240PYR$_{14}$TFSI & 423 & 70\\
    &&90LiTFSI/210PYR$_{14}$TFSI & 423 & 70\\
    \hline
    \multirow{7}{*}{WMI-MD} & \multirow{7}{*}{APPLE\&P} & 200PYR$_{14}$BF$_4$ & 423 & 102\\
    &&200PYR$_{14}$TFSI & 423 & 43\\
    &&20LiTFSI/180PYR$_{14}$TFSI & 423 & 42\\
    &&20LiTFSI/180PYR$_{14}$TFSI & 353 & 110\\
    &&20LiTFSI/180PYR$_{14}$TFSI & 323 & 80\\
    &&40LiTFSI/160PYR$_{14}$TFSI & 423 & 66\\
    &&60LiTFSI/140PYR$_{14}$TFSI & 423 & 42\\
    \hline
    DPMD & PBE & 300PYR$_{14}$BF$_4$ & 423 & 20\\
    \hline
    \multirow{7}{*}{MACE} & \multirow{7}{*}{PBE} & 100PYR$_{14}$BF$_4$ & 423 & 7.5\\
    &&100PYR$_{14}$TFSI & 423 & 15.5\\
    &&10LiTFSI/90PYR$_{14}$TFSI & 423 & 13.5\\
    &&10LiTFSI/90PYR$_{14}$TFSI & 353 & 10.5\\
    &&10LiTFSI/90PYR$_{14}$TFSI & 323 & 9\\
    &&20LiTFSI/80PYR$_{14}$TFSI & 423 & 10\\
    &&30LiTFSI/70PYR$_{14}$TFSI & 423 & 10.5\\
    \hline
  \end{tabular}
\end{table}

\begin{table}[p]
\centering
  \caption{Information of the AIMD dataset and the LAMMPS training initial dataset}
  \label{tbl:AIMDdataset}
  \begin{tabular}{cccccc}
    \hline
    & \multirow{2}{*}{Force Field} & \multirow{2}{*}{Component} & \multirow{2}{*}{Eq?} & Number of & \multirow{2}{*}{etc.} \\
    &&&& frames \\
    \hline
    \multirow{3}{*}{AIMD dataset} & \multirow{3}{*}{PBE} & \multirow{3}{*}{300PYR$_{14}$BF$_4$} & yes & 4000 & box 14 $\sim$ 16 \AA \\
    &&& yes & 1000 & box 15.5 $\sim$ 16.5 \AA \\
    &&& no & 6000 & box 15.0 $\sim$ 16.2 \AA \\ 
    \hline
    & \multirow{2}{*}{Force Field} & Component & \multirow{2}{*}{Eq?} & Number of & \multirow{2}{*}{Added frames} \\
    && (Total DPGEN iterations) && initials \\
    \hline \\
    \multirow{1}{*}{Training initials} & \multirow{1}{*}{PBE} &  300PYR$_{14}$BF$_4$ (9) & no & 250 & 19812 \\
    \hline
  \end{tabular}
\end{table}

\begin{table}[p]
\centering
  \caption{Information of the validation AIMD dataset.}
  \label{tbl:validation}
  \begin{tabular}{ccccc}
    \hline
    \multirow{2}{*}{Functional} & \multirow{2}{*}{Component} & \multirow{2}{*}{Eq?} & Number of & Number of \\
    &&&initials&frames \\
    \hline
    \multirow{4}{*}{PBE} & \multirow{2}{*}{10PYR$_{14}$BF$_4$} & yes & 200 & 1 \\
    &&no&200&1\\ \cline{2-5}
    & \multirow{2}{*}{10PYR$_{14}$TFSI} & yes & 200 & 1 \\
    &&no&200&1\\ \cline{2-5}
    & \multirow{2}{*}{1LiTFSI/9PYR$_{14}$TFSI} & yes & 200 & 1 \\
    &&no&200&1\\ \cline{2-5}
    & \multirow{2}{*}{2LiTFSI/8PYR$_{14}$TFSI} & yes & 200 & 1 \\
    &&no&200&1\\ \cline{2-5}
    & \multirow{2}{*}{3LiTFSI/7PYR$_{14}$TFSI} & yes & 200 & 1 \\
    &&no&200&1\\ \cline{2-5}
    \hline
    \multicolumn{5}{l}{*200 frames for each system}\\
    \multicolumn{5}{l}{*All NVT, box size -1\AA ~+1\AA\space from EQ box size}
  \end{tabular}
\end{table}

\begin{figure}[p]
\centering
  \includegraphics[height=15cm]{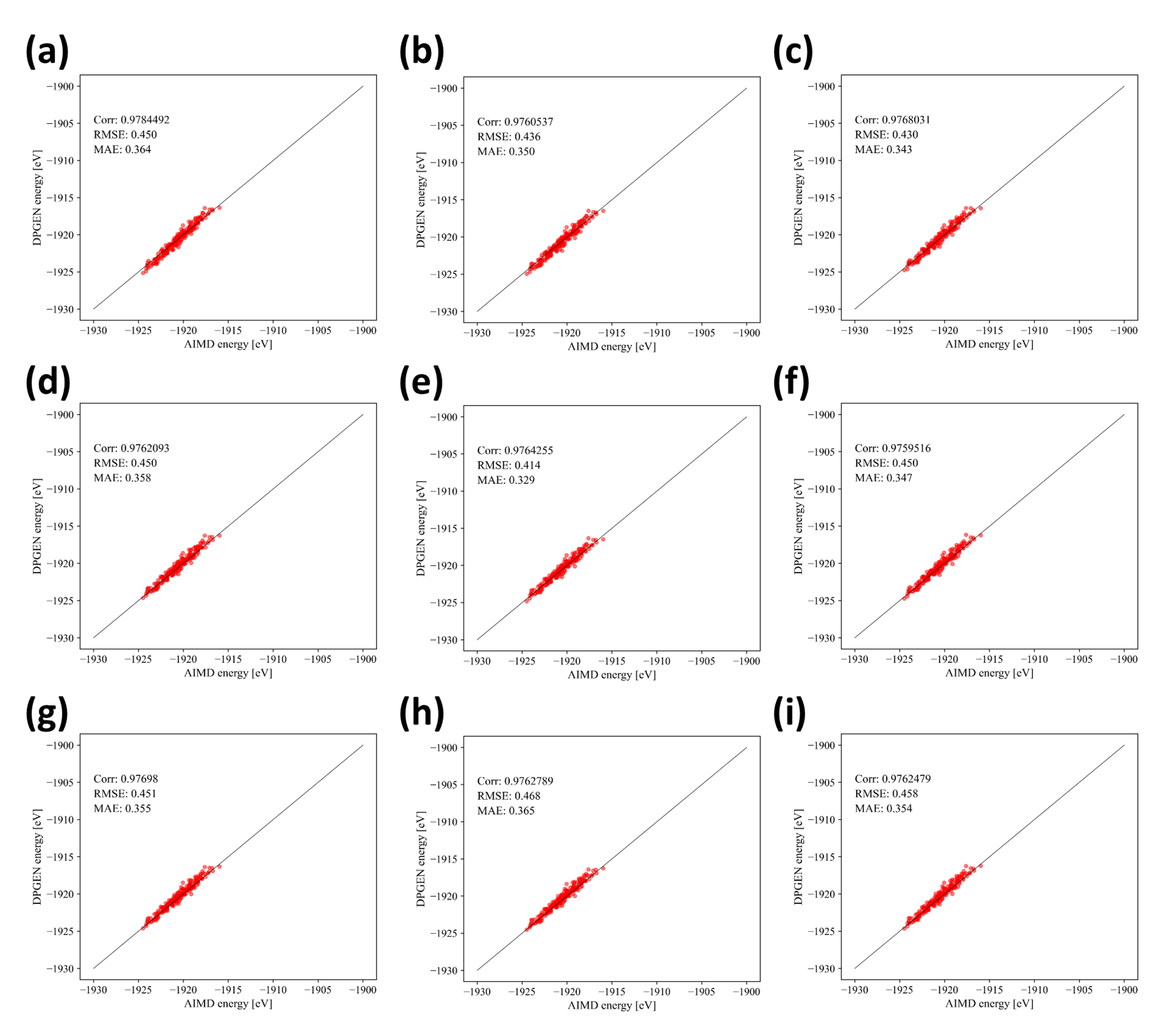}
  \caption{Energy validation of each DPGEN MLFF iteration compared to AIMD calculation, total 9 iterations from (a) to (i). Validation test consist of 200 EQ structures, where each structure consists of 10 pairs of BF$_4$PYR$_{14}$.}
  \label{fgr:validEQ_E}
\end{figure}
\begin{figure}[p]
\centering
  \includegraphics[height=15cm]{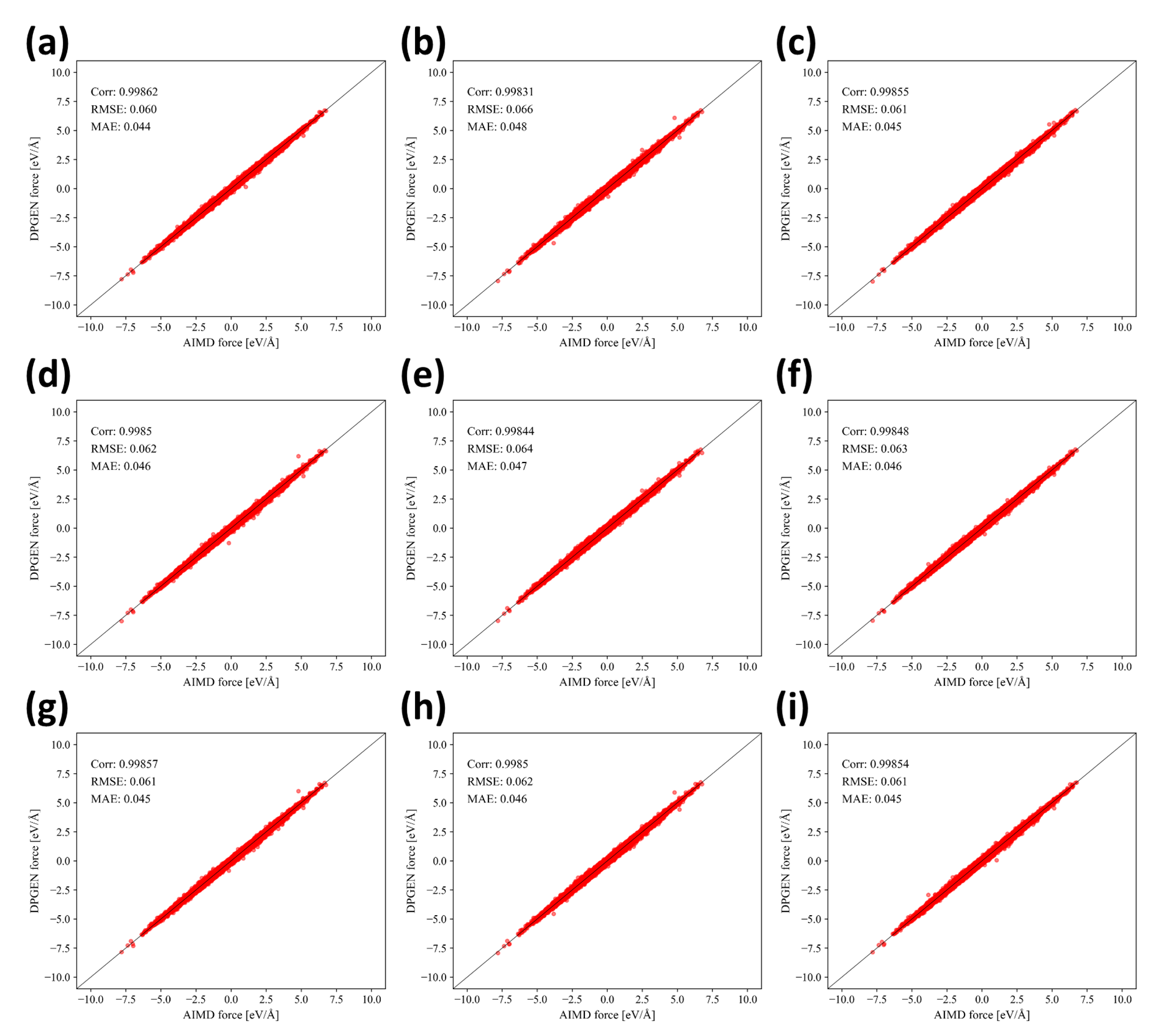}
  \caption{Force validation of each DPGEN MLFF iteration compared to AIMD calculation, total 9 iterations from (a) to (i). Validation test consist of 200 EQ structures, where each structure consists of 10 pairs of BF$_4$PYR$_{14}$.}
  \label{fgr:validEQ_F}
\end{figure}
\begin{figure}[p]
\centering
  \includegraphics[height=15cm]{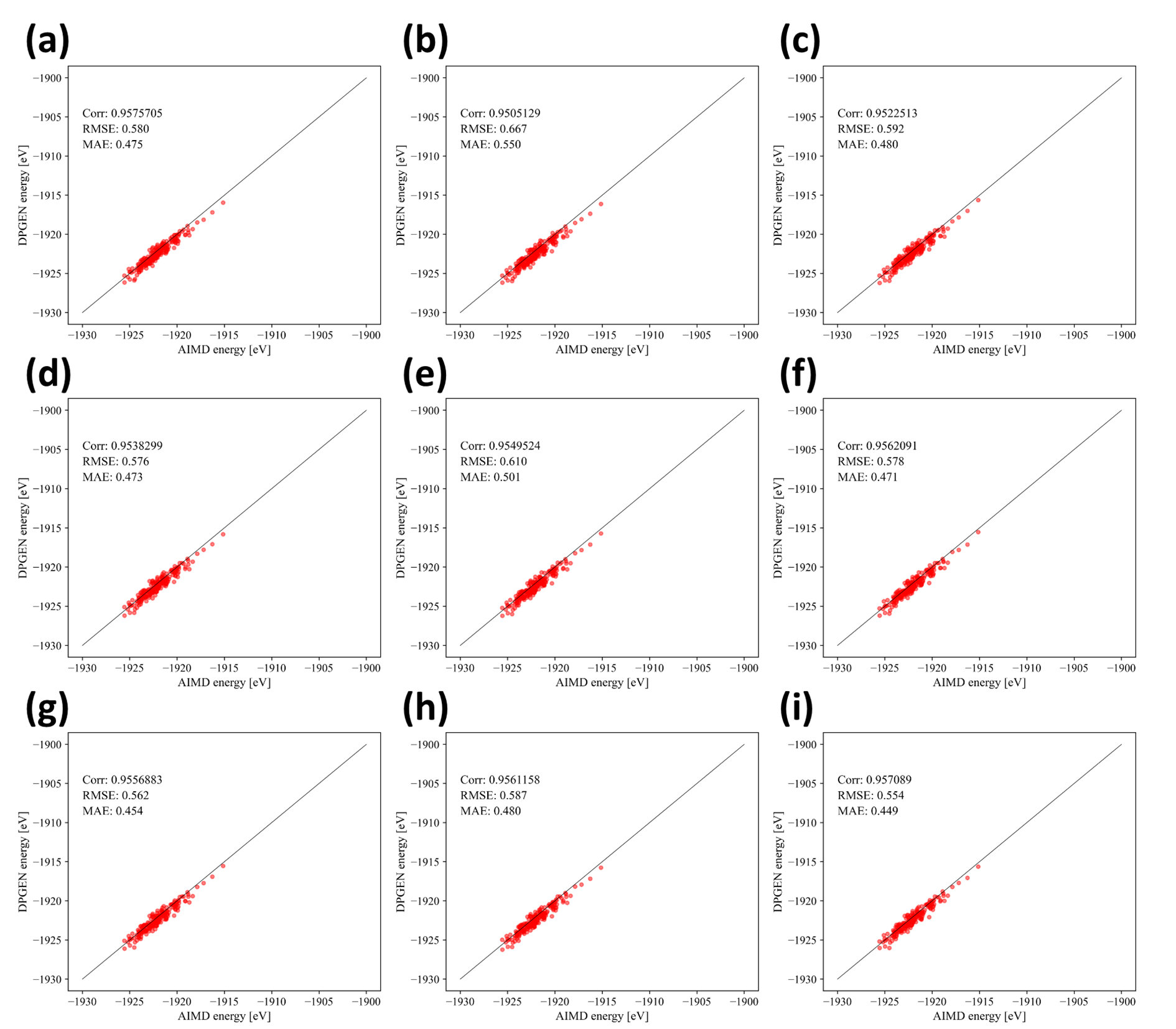}
  \caption{Energy validation of each DPGEN MLFF iteration compared to AIMD calculation, total 9 iterations from (a) to (i). Validation test consist of 200 nEQ structures, where each structure consists of 10 pairs of BF$_4$PYR$_{14}$.}
  \label{fgr:validnEQ_E}
\end{figure}
\begin{figure}[p]
\centering
  \includegraphics[height=15cm]{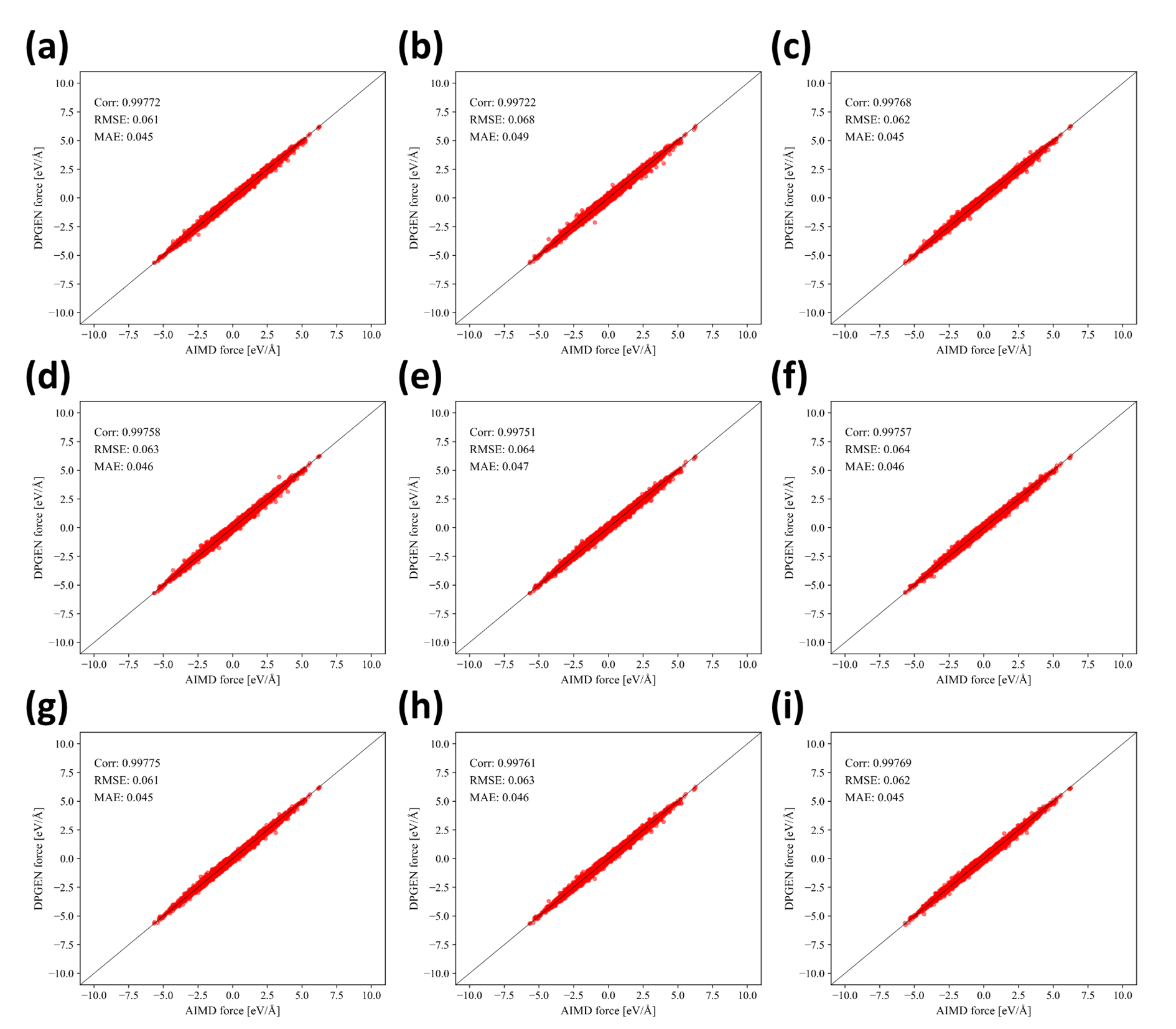}
  \caption{Force validation of each DPGEN MLFF iteration compared to AIMD calculation, total 9 iterations from (a) to (i). Validation test consist of 200 nEQ structures, where each structure consists of 10 pairs of BF$_4$PYR$_{14}$.}
  \label{fgr:validnEQ_F}
\end{figure}

\begin{figure}[p]
\centering
  \includegraphics[height=7cm]{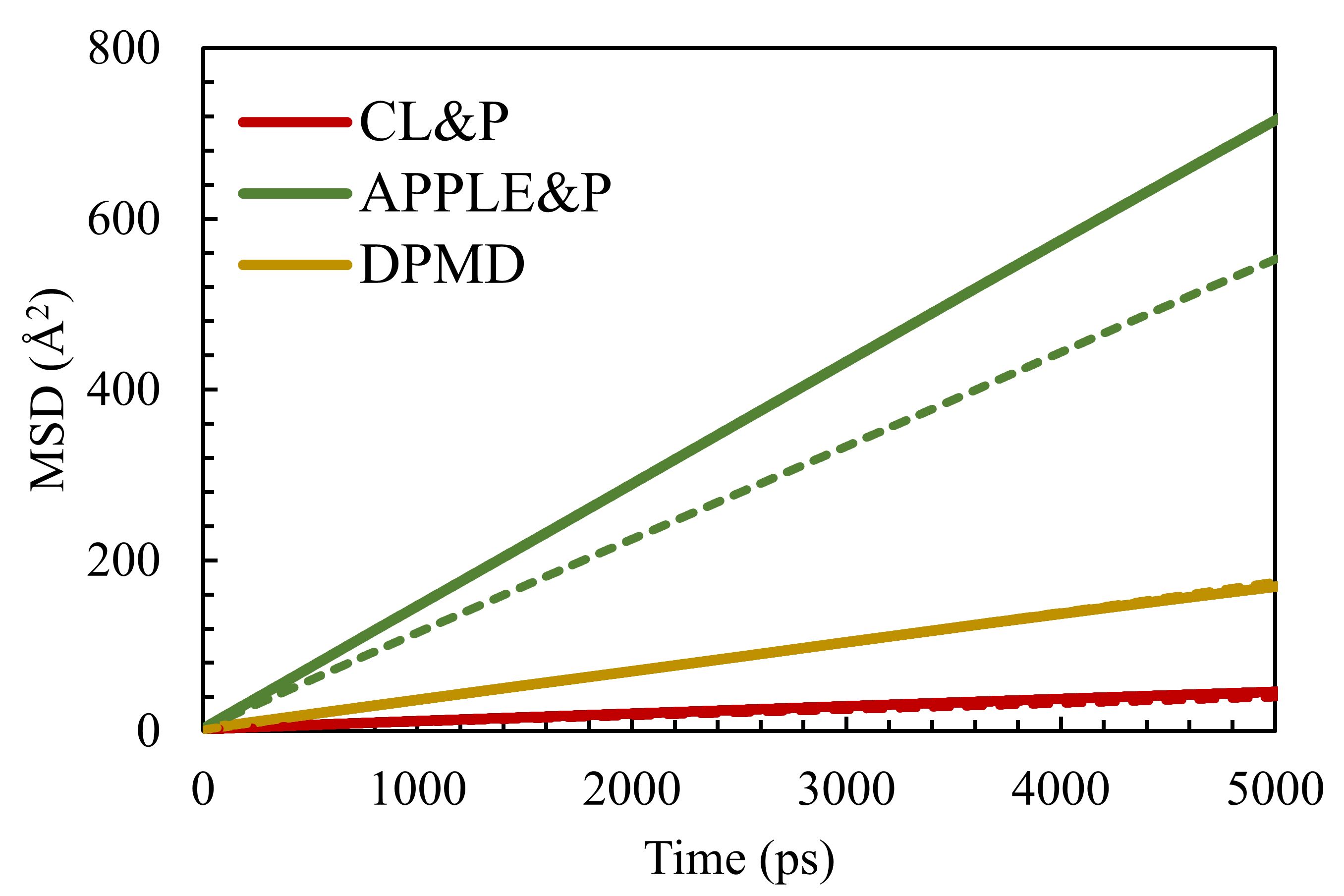}
  \caption{MSD plot calculated from CL\&P, APPLE\&P and DPMD MD simulation. MSD of anion BF$_4$ is represented as solid line and MSD of cation PYR$_{14}$ is represented as dashed line.}
  \label{fgr:DPGEN_MSD}
\end{figure}

\begin{figure}[p]
\centering
  \includegraphics[height=9cm]{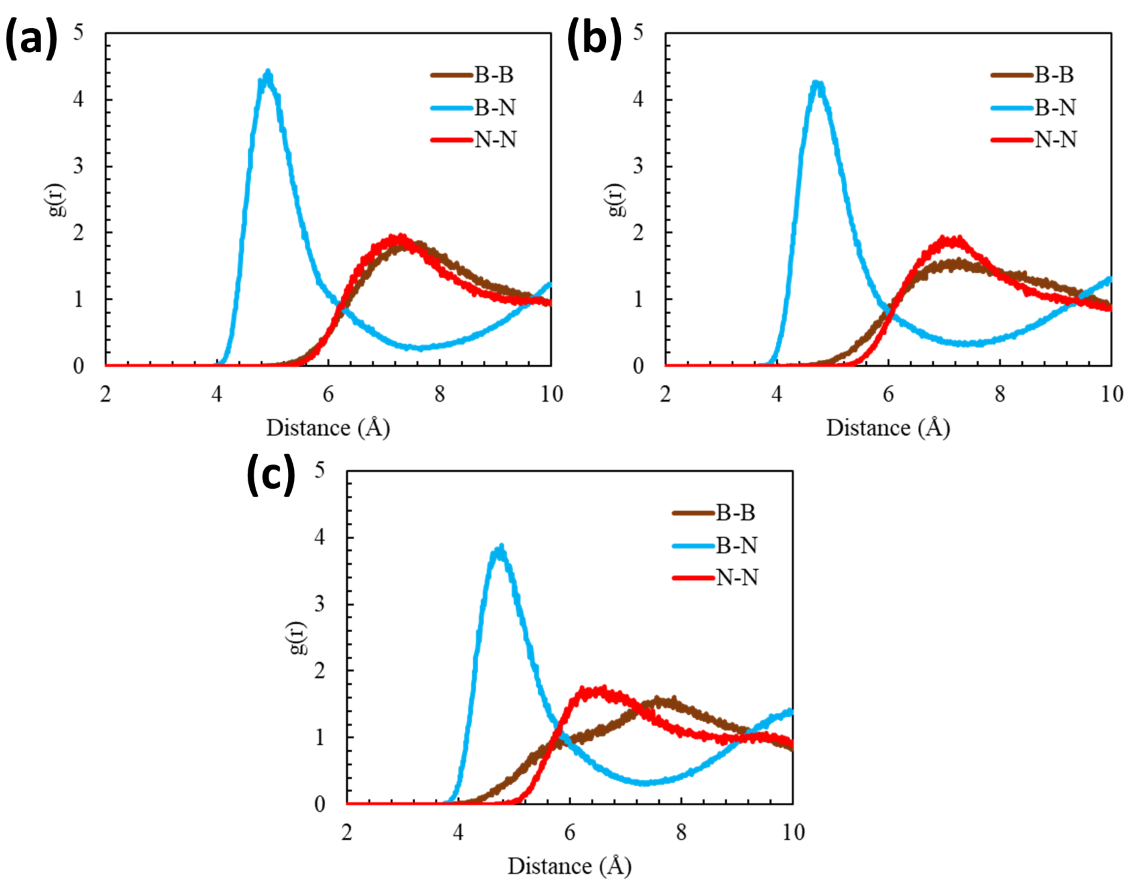}
  \caption{Radial distribution function of B(BF$_4$)-B, B-N(PYR$_{14}$), and N-N calculated from (a) CL\&P, (b) APPLE\&P, and (c) DPMD MD simulation.}
  \label{fgr:BF4PYR14_RDF}
\end{figure}

\begin{figure}[p]
\centering
  \includegraphics[height=6cm]{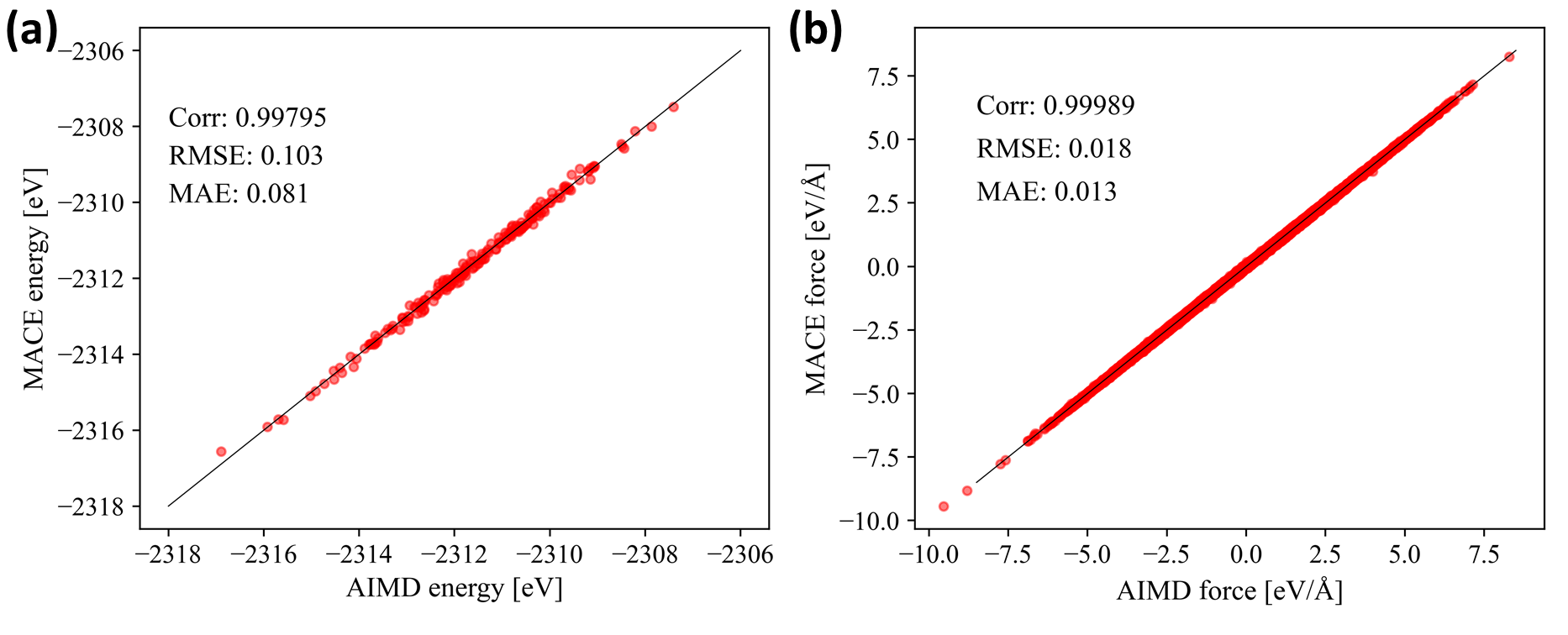}
  \caption{(a) Energy and (b) force validation test of MACE MLFF with LiTFSI 10\% LiTFSI/TFSIPYR$_{14}$ IL.}
  \label{fgr:trainLi1}
\end{figure}
\begin{figure}[p]
\centering
  \includegraphics[height=6cm]{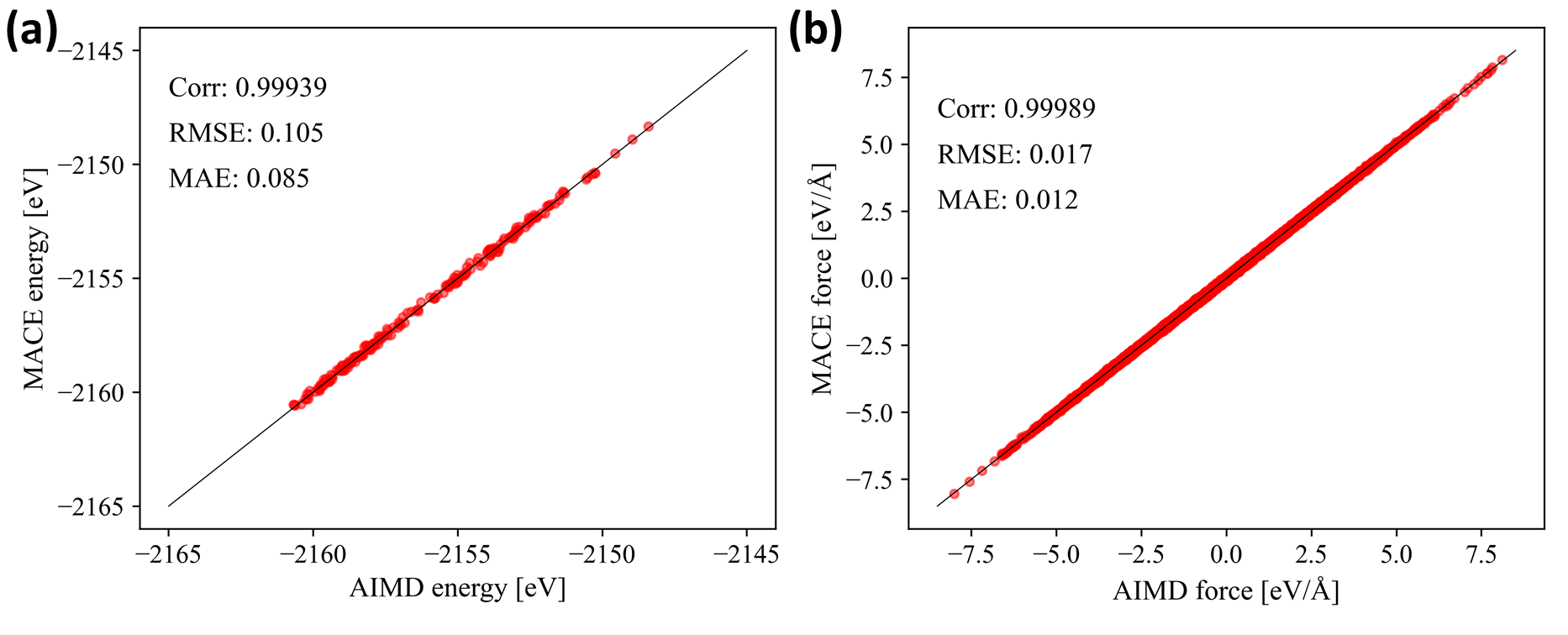}
  \caption{(a) Energy and (b) force validation test of MACE MLFF with LiTFSI 20\% LiTFSI/TFSIPYR$_{14}$ IL.}
  \label{fgr:trainLi2}
\end{figure}
\begin{figure}[p]
\centering
  \includegraphics[height=6cm]{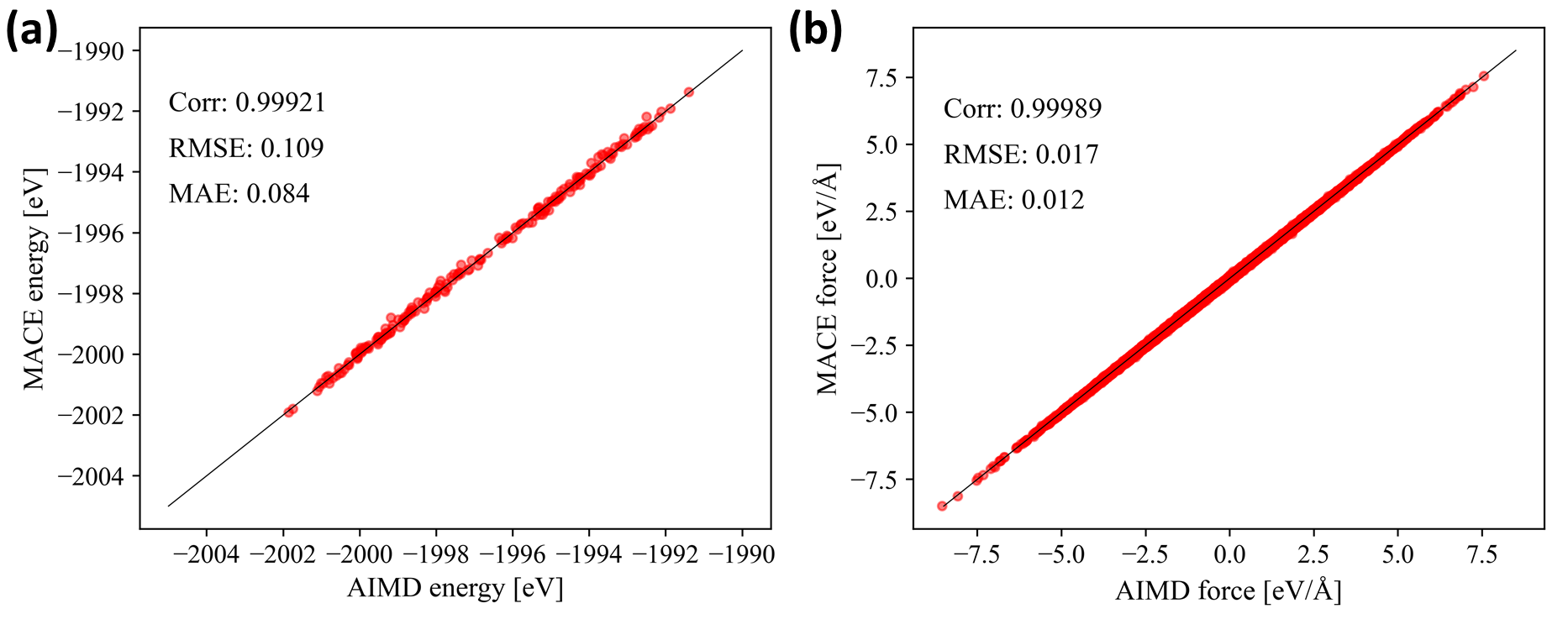}
  \caption{(a) Energy and (b) force validation test of MACE MLFF with LiTFSI 30\% LiTFSI/TFSIPYR$_{14}$ IL.}
  \label{fgr:trainLi3}
\end{figure}
\begin{figure}[h]
\centering
  \includegraphics[height=6cm]{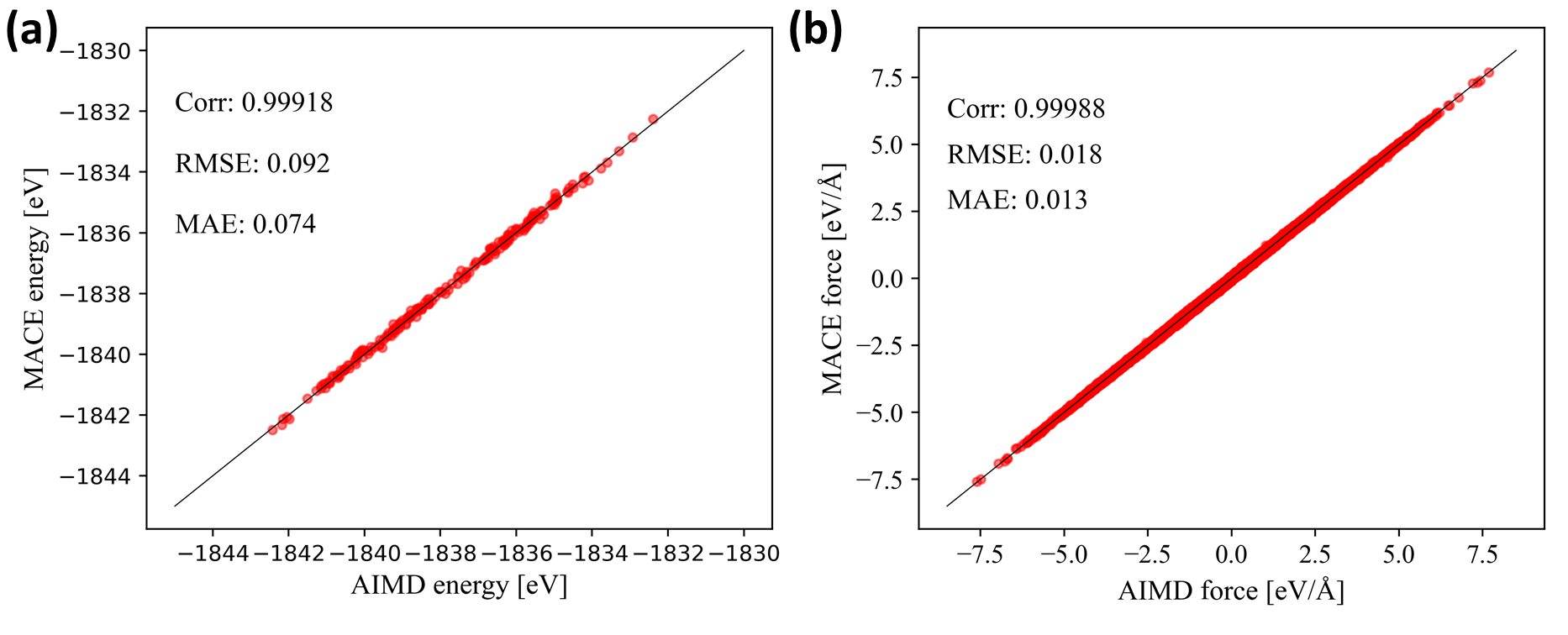}
  \caption{(a) Energy and (b) force validation test of MACE MLFF with LiTFSI 40\% LiTFSI/TFSIPYR$_{14}$ IL.}
  \label{fgr:trainLi4}
\end{figure}

\begin{figure}[p]
\centering
  \includegraphics[height=16cm]{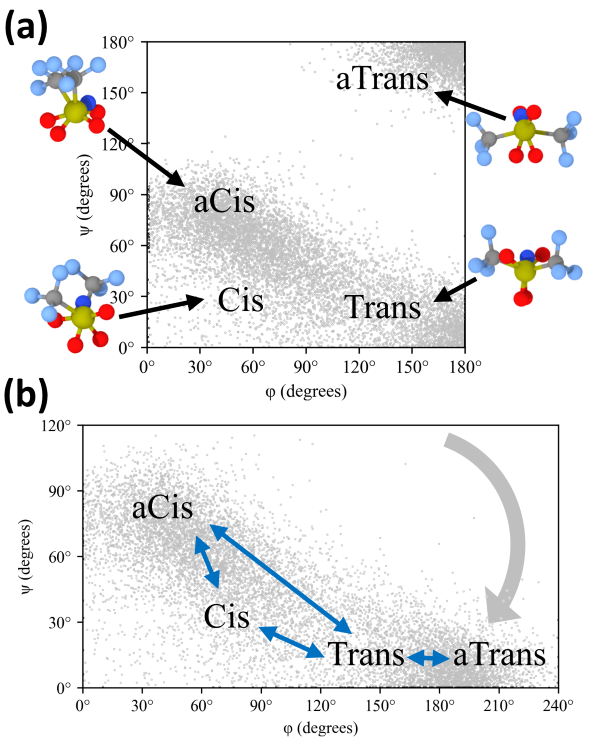}
  \caption{(a) Dot distribution plot of a single TFSI in vacuum environment run by WMI-MD APPLE\&P force field (Pol) classified by 4 structures $Cis$, $aCis$, $Trans$, and $aTrans$ and (b) example of a TFSI transition from $aCis$ to $aTrans$}
  \label{fgr:tfsi_detail}
\end{figure}
\begin{figure}[p]
\centering
  \includegraphics[height=10cm]{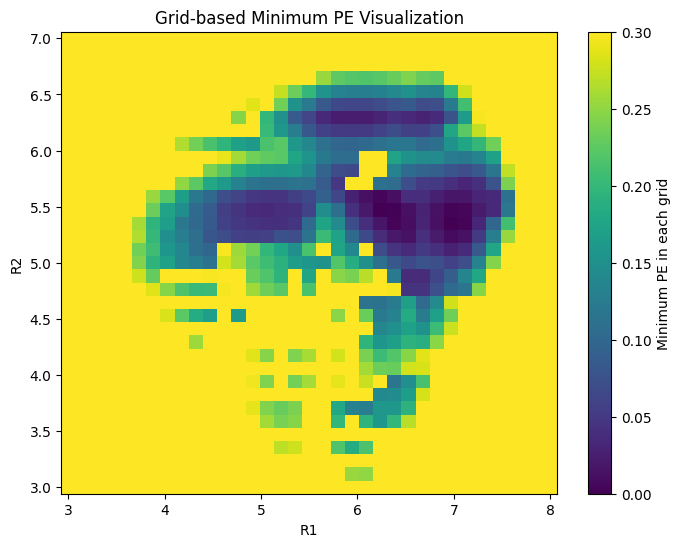}
  \caption{PES of PYR$_{14}$ calculated from DFT simulation with B3LYP functional.}
  \label{fgr:PYR_PES}
\end{figure}

\bibliography{refs}